\documentclass[fleqn,usenatbib]{mnras}

\usepackage{newtxtext,newtxmath}

\usepackage[T1]{fontenc}
\usepackage{ae,aecompl}

\usepackage{graphicx}	
\usepackage{amsmath}	
\usepackage{amssymb}	

\newcommand\<{\langle}
\renewcommand\>{\rangle}

\newcommand\bM{\mathbf M}

\newcommand\bS{\mathbf S}
\newcommand\bbrho{\mathbf \rho}
\newcommand\rg{GM/c^2}

\newcommand\aI{\alpha_I}
\newcommand\aQ{\alpha_Q}
\newcommand\aU{\alpha_U}
\newcommand\aV{\alpha_V}
\newcommand\aP{\alpha_P}
\newcommand\chaPs{\cosh(\alpha_P \lambda)}
\newcommand\shaPs{\sinh(\alpha_P \lambda)}
\newcommand\jI{j_I}
\newcommand\jQ{j_Q}
\newcommand\jU{j_U}
\newcommand\jV{j_V}
\newcommand\eaIs{e^{-\alpha_I \lambda}}
\newcommand\eaPs{e^{-\alpha_P \lambda}}
\newcommand\acj{\alpha \cdot j}
\newcommand\acbS{\alpha \cdot \bS}

\newcommand\LDI{$\rm LDI^2$}

\title[ipole - semianalytic scheme for polarized transport]{ipole - semianalytic
  scheme for relativistic polarized radiative transport}

\author[M. Mo{\'s}cibrodzka and C.F. Gammie]{
M. Mo{\'s}cibrodzka$^{1}$\thanks{E-mail: m.moscibrodzka@astro.ru.nl}, 
C.F. Gammie$^{2}$\thanks{E-mail: gammie@illinois.edu}
\\
$^{1}$Department of Astrophysics/IMAPP, Radboud University, P.O. Box 9010, 6500 GL Nijmegen, The Netherlands\\
$^{2}$Department of Astronomy and Department of Physics, University of Illinois, Urbana, IL 61801, USA
}

\date{Accepted XXX. Received YYY; in original form ZZZ}

\pubyear{2015}

\begin{document}
\label{firstpage}
\pagerange{\pageref{firstpage}--\pageref{lastpage}}
\maketitle

\begin{abstract}

We describe {\tt ipole}, a new public ray-tracing code for covariant,
polarized radiative transport.  The code extends the {\tt ibothros} scheme for
covariant, unpolarized transport using two representations of the polarized
radiation field: in the coordinate frame, it parallel transports the coherency
tensor; in the frame of the plasma it evolves the Stokes parameters under
emission, absorption, and Faraday conversion. The transport step is
implemented to be as spacetime- and coordinate- independent as possible. The
emission, absorption, and Faraday conversion step is implemented using an
analytic solution to the polarized transport equation with constant
coefficients.  As a result, {\tt ipole} is stable, efficient, and produces a physically
reasonable solution even for a step with high optical depth and Faraday depth.
We show that the code matches analytic results in flat space, and that it
produces results that converge to those produced by Dexter's {\tt grtrans}
polarized transport code on a complicated model problem.  We expect {\tt
  ipole} will mainly find applications in modeling Event Horizon Telescope
sources, but it may also be useful in other relativistic transport problems
such as modeling for the IXPE mission.
\end{abstract}




\section{Introduction}

The Event Horizon Telescope (EHT) will soon produce full polarization images
of the luminous plasma surrounding the event horizon in the low accretion rate
systems Sgr A* and M87* \citep{johnson:2015}. Much of the information content of EHT
observations will be in the polarized components of the radiation field;
extracting this information will require a model for the state of the
radiating plasma as well as the ability to produce mock full polarization
observations of these models. Although mock total intensity observations of
accretion flow and jet models have now become common
\citep{falcke:2000a,noble:2007,broderick:2009a,broderick:2009b,dexter:2009,moscibrodzka:2009,yuan:2009,dexter:2010,broderick:2011a,broderick:2011b,dexter:2011,vincent:2011,dexter:2012,dolence:2012,moscibrodzka:2012,younsi:2012,dexter:2013,chan:2013,moscibrodzka:2014,chan:2015,vincent:2015,younsi:2015,ball:2016,fraga:2016,moscibrodzka:2016,pu:2016a,pu:2016b,chan:2017,mao:2017,medeiros:2017,porth:2017,shiokawa:2017,roelofs:2017}, full polarization models - although not completely novel
\citep{bromley:2001,bb:2004,broderick:2005,huang:2008,romans:2012,gold:2016,moscibrodzka:2017} - are less well explored.

In EHT target models millimeter photons are produced by synchrotron emission.
It is therefore natural that EHT targets have substantial linear polarization,
and indeed the linear polarization fraction in Sgr A* is $\sim 7$ per
cent\footnote{It is worth mentioning that NIR emission from Sgr A* also has strong
linear polarization, of 20-40 per cent \citep{eckart:2008,sha:2015}.} 
\citep{bower:2003,bower:2005,marrone:2007,marrone:2008}
and in M87 it is $ < 1$ per cent \citep{kuo:2014}.
Circular polarization can also be produced in emission and by
Faraday conversion of linearly polarized radiation.  The circular polarization
fraction in Sgr A* has been measured as 1.2-1.6 per cent \citep{munoz:2009,munoz:2012}. 
Our interest in polarized models is therefore well motivated.

Total intensity models of accreting black holes manifest familiar relativistic
effects \citep{cunningham:1973,cunningham:1975}: gravitational lensing, doppler shift, doppler boosting, and
gravitational redshift all contribute at order unity to models of accretion
flows where the bulk of the emission is generated close to the event horizon.
To this, full polarization models add ``gravitational Faraday rotation''
\citep{balazs:1958}, i.e. the spacetime can rotate the plane of polarization of
an electromagnetic wave.  In the weak field limit, the rotational angle is
proportional to the line of sight component of the angular momentum of the
lensing mass \citep{ishihara:1988}.

Several existing codes are capable of generating polarized images of radiating
plasma near a compact object. \footnote{Polarized transport schemes already
  exist for applications in cosmology, but typically do not use ray-tracing.}
Of these, only Dexter's {\tt grtrans} code \citep{dexter:2016} has been
publicly released.  It seems to us that it is useful to have multiple,
distinct, publicly available solutions of the problem, for verification
purposes.  Nevertheless, our code is not completely independent and owes much
to the careful testing and thoughtful construction of {\tt grtrans}.

Still, our scheme differs from {\tt grtrans} in three significant respects.

First, in the formulation of the Liouville operator (the convective derivative
operator in phase space): we use parallel transport of a coherency or photon
density tensor rather than direct integration of the invariant Stokes
parameters with a rotation term for linear polarization.  The coherency matrix
approach, analogous to that developed by \cite{vanball:1985}, seems
conceptually cleaner to us and requires relatively little thought (and
therefore reduces the scope for error in, for example, formulating a
polarization measurement).  It is also manifestly covariant, so it is easy to
change coordinate systems.

Second, at each step we use an analytic solution for polarized
transport with constant absorption, emission, and rotation coefficients
(defined below).   The solution was first written down by \cite{LDI:1985}.  We
recount it below, as well as a few special cases in an appendix.   
The result is a cheap second-order scheme that behaves well
even when the absorption optical depth and/or Faraday depth is large over a
single step. 

Third, we directly integrate the geodesic equation rather than using {\tt
  geokerr} \citep{dexter:2009}, which relies on integrability of geodesics in the Kerr metric.
Again, this makes our code coordinate and spacetime independent.  We can
therefore study polarization properties of non-GR black hole models, and
switch to unconventional coordinate systems (such as the Cartesian Kerr-Schild
coordinates used by, e.g., BHAC code, \citealt{porth:2017}) for the geodesic integration.

In the end, the value of each of these differences is somewhat subjective.
What is not subjective is the value of having quasi-independent schemes for
solving a complicated, technically demanding problem like relativistic
polarized radiative transport.

This paper is organized as follows. In section~\ref{sect:equations} we present
the equations of polarized radiative transfer through magnetized plasma.
Section~\ref{sect:equations} outlines the coherency tensor formalism of
\citet{gammie:2012}, however we also clarify a few points from that
paper.  In section~\ref{sect:numerical}, we describe a semi-analytic scheme
for solving the equations in arbitrary geometry.  In section~\ref{sect:test},
we present a few simple tests and demonstrate the performance of the numerical
scheme in recovering known analytic solutions of the polarized transfer
equations.  In case of more complex problems, that do not have analytic
solutions, we compare {\tt ipole} numerical results to the results obtained
with {\tt grtrans}. We summarize the paper and conclude in
section~\ref{sect:summary}.

\section{Governing Equations}\label{sect:equations}

The radiative transfer equation for time-independent, unpolarized, nonrelativistic transport, including emission and absorption but not scattering, is
\begin{equation}
\frac{d I_\nu}{ds} = j_\nu - \alpha_\nu I_\nu,
\end{equation}
where $I_\nu \equiv$ specific intensity, $\nu \equiv frequency$, $j_\nu \equiv$ emissivity, and $\alpha_\nu \equiv$ absorptivity.  Each term is frame dependent.  The covariant generalization is
\begin{equation}\label{covunpol}
\frac{d}{d\lambda}\left(\frac{I_\nu}{\nu^3}\right) 
= \left(\frac{j_\nu}{\nu^2}\right) - \left(\nu \alpha_\nu\right)
\left(\frac{I_\nu}{\nu^3}\right),
\end{equation}
where $\lambda \equiv$ the affine parameter along a photon trajectory, $d/d\lambda$ is the convective derivative in phase space (``Liouville operator''), and each term in parentheses is invariant and can thus be evaluated in any frame.
The affine parameter is defined through the geodesic equations
\begin{equation}
\frac{d x^\mu}{d\lambda} = k^\mu
\end{equation}
and
\begin{equation}
\frac{d k^\mu}{d\lambda} = -\Gamma^\mu_{\alpha\beta} k^\alpha k^\beta,
\end{equation}
where $k^\mu \equiv$ wave four-vector and $\Gamma \equiv$ connection coefficients.  The frequency measured by an observer with four-velocity $u^\mu$ is
\begin{equation}
\omega = -k^\mu u_\mu.
\end{equation}
The relationship between $\omega$ and the frequency in Hz measured by the observer depends on the units of $k^\mu$.  We have implicitly assumed (and will continue to assume below) that photons travel along null geodesics and therefore that $\nu$ is large compared to the plasma frequency and electron gyrofrequency \citep[see][for a more general treatment]{bb:2004}.  In EHT sources this is an excellent approximation. 

The radiative transfer equation for {\em polarized}, time-independent, nonrelativistic transport, including emission and absorption but not scattering, is
\begin{equation}\label{poltrans}
\frac{d}{ds}
\begin{pmatrix} I_\nu \\ Q_\nu \\ U_\nu \\ V_\nu \end{pmatrix}
 = \begin{pmatrix} j_{\nu,I} \\ j_{\nu,Q} \\ j_{\nu,U} \\ j_{\nu,V} \end{pmatrix}
- \begin{pmatrix} 
\alpha_{\nu,I} & \alpha_{\nu,Q} & \alpha_{\nu,U} & \alpha_{\nu,V} \\
\alpha_{\nu,Q} & \alpha_{\nu,I} & \rho_{\nu,V} & -\rho_{\nu,U} \\
\alpha_{\nu,U} & -\rho_{\nu,V} & \alpha_{\nu,I} & \rho_{\nu,Q} \\
\alpha_{\nu,V} & \rho_{\nu,U} & -\rho_{\nu,Q} & \alpha_{\nu,I}  
\end{pmatrix}
\begin{pmatrix} I_\nu \\ Q_\nu \\ U_\nu \\ V_\nu \end{pmatrix}, 
\end{equation}
where $I_\nu, Q_\nu, U_\nu, V_\nu$ are (frame-dependent)
specific intensities associated with the Stokes parameters.\footnote{The sign of $\rho_U$ differs from \citet{dexter:2016} and agrees with \citet{LDI:1985}, but this has no effect on the \citet{dexter:2016} solution because $\rho_U = 0$ in the frame in which the transfer coefficients are evaluated.}  Notice that $Q_\nu, U_\nu, V_\nu$ are signed quantities while $I_\nu$ is positive definite.  $Q_\nu > 0$ 
corresponds to linear polarization along one axis in the
plane perpendicular to the wave 3-vector, while $Q_\nu < 0$ 
corresponds to linear polarization along the second axis.  
$U_\nu$ describes polarization at $\pm 45$deg to the first axis. 
$V_\nu$ is circular polarization. Positive $V_\nu$ always means right-hand circular  polarization (RCP). The IEEE convention is that for RCP the electric field vector rotates in a right-handed direction at a fixed position if thumb points along wavevector $k^\mu$.  For RCP the field rotates counter-clockwise as seen from the observer \citep[see][for a discussion]{hamaker:1996}

Equation (\ref{poltrans}) has $11$ transfer coefficients that depend on physical conditions in the plasma.  These are the four emission coefficients $j_{\nu,A}$ (subscript $A$ can be one of $I,Q,U,V$); the four absorption coefficients $\alpha_{\nu,A}$, and the three rotation coefficients $\rho_{\nu,A}$.  By definition $I_\nu^2 \ge Q_\nu^2 + U_\nu^2 + V_\nu^2$, i.e. the polarization fraction is $\le 100\%$, and evidently we must have
\begin{equation}
j_{\nu,I}^2 > j_{\nu,Q}^2 + j_{\nu,U}^2 + j_{\nu,V}^2
\end{equation}
to guarantee this.  Notice that $j_{\nu,I} > 0$, but $j_{\nu,Q}, j_{\nu,U}, j_{\nu,V}$ can have either sign.  Assuming maser action is absent, $\alpha_{\nu,I} > 0$, but $\alpha_{\nu,Q}, \alpha_{\nu,U}, \alpha_{\nu,V}$ can also have either sign.

The covariant generalization of (\ref{poltrans}) is not as simple as for the unpolarized transfer equation because the definition of $Q_\nu, U_\nu$ depend on the orientation of the axes by the observer who makes the measurement. \citet{bb:2004} have presented a generalization of (\ref{covunpol}) in terms of the ``invariant'' Stokes parameters $\bS \equiv (I,Q,U,V) \equiv (I_\nu, Q_\nu, U_\nu, V_\nu)/\nu^3$ that explicitly accounts for the rotation of an observer frame along the line of sight (in our notation, the absence of subscript $\nu$ implies an invariant quantity; thus $\alpha_I \equiv \nu \alpha_{\nu,I}$). This generalization has been used by \citet{broderickloeb:2009}, \citet{romans:2012}, \citet{gold:2016}, \citet{dexter:2016}, \citet{moscibrodzka:2017} to generate polarized models of accretion onto a black hole.

The covariant Stokes formulation of the polarized transfer equation is not
written in manifestly covariant form, and hence the transformation of Stokes
parameters from one frame to another is not completely transparent, although
in the end, it amounts to a rotation.  \citet{gammie:2012} (see also
\citealt{kosowsky:1996}, \citealt{weinberg:2008}) rewrote the polarized transport equation in terms of the rank-2, Hermitian, coherency tensor
\begin{equation}
N^{\alpha\beta} \equiv C\, \< a_{k}^\alpha a^{*\beta}_k \>,
\end{equation}
where $a_k$ is a Fourier component of the four-vector potential and $C$ is an arbitrary constant.  This description is manifestly covariant.

Let us relate $N^{\alpha\beta}$ to the Stokes parameters defined in an orthonormal tetrad $e_{(a)}^\mu$ (parenthesized lowercase roman letters indicate tetrad indices).  We make two assumptions about the tetrad: $e_{(t)}^\mu = u^\mu$, the four-velocity of the associated observer; and $e_{(3)}^\mu = k^\mu - \omega u^\mu$.  In words: the third spatial basis element is a unit vector oriented parallel to the spatial component of the wavevector.  

It is then helpful to define four auxiliary tensors in the tetrad
frame:
\begin{equation}
m_I \equiv
\begin{pmatrix} 
0 & 0 & 0 & 0 \\
0 & 1 & 0 & 0 \\
0 & 0 & 1 & 0 \\
0 & 0 & 0 & 0  
\end{pmatrix},
\end{equation}
\begin{equation}
m_Q \equiv 
\begin{pmatrix} 
0 & 0 & 0 & 0 \\
0 & 1 & 0 & 0 \\
0 & 0 & -1 & 0 \\
0 & 0 & 0 & 0  
\end{pmatrix},
\end{equation}
\begin{equation}
m_U \equiv 
\begin{pmatrix} 
0 & 0 & 0 & 0 \\
0 & 0 & 1 & 0 \\
0 & 1 & 0 & 0 \\
0 & 0 & 0 & 0  
\end{pmatrix},
\end{equation}
\begin{equation}
m_V \equiv 
\begin{pmatrix} 
0 & 0 & 0 & 0 \\
0 & 0 & -i & 0 \\
0 & i & 0 & 0 \\
0 & 0 & 0 & 0  
\end{pmatrix}.
\end{equation}
These are just the Pauli matrices (see \citealt{lopezariste:1999} for a
discussion) 
in the two dimensional space perpendicular to $u^\mu$ and the wave three-vector.  Then we define $C$ so that
\begin{equation}\label{eq:StoN}
N^{(a)(b)} = m^{(a)(b)}_A S_A
\end{equation}
(again, the index $A$ is one of $I,Q,U,V$), $S_A$ is a component of the invariant Stokes vector $\bS$, and summation over $A$ is implied. The inverse relation is 
\begin{equation}\label{eq:NtoS}
S_A = \frac{1}{2} m^{*(a)(b)}_A N_{(a)(b)}.
\end{equation}
These linear relations between $N$ and $S$ are easy to implement numerically.  It is also obvious how $N$ transforms under boosts, rotations, and general coordinate transformations, because it is a tensor.

The covariant polarized transport equation is 
\begin{equation}\label{eq:covtrans}
k^\mu \nabla_\mu N^{\alpha\beta} = J^{\alpha\beta} + H^{\alpha\beta\gamma\delta}
	N_{\gamma\delta}.
\end{equation}
Here $\nabla_\mu$ is a covariant derivative (the derivative operator is understood to follow a photon trajectory in frequency space), $J^{\alpha\beta}$ is an emissivity tensor, and $H^{\alpha\beta\gamma\delta}$ incorporates absorption and Faraday rotation.  Expanding the covariant derivative in a coordinate basis, (\ref{eq:covtrans}) becomes
\begin{equation}\label{eq:covtrans2}
\frac{d N^{\alpha\beta}}{d\lambda} 
= 
- \Gamma^\alpha_{\mu\nu} k^\mu N^{\nu\beta}
- \Gamma^\beta_{\mu\nu} k^\mu N^{\alpha\nu}
+ J^{\alpha\beta} + H^{\alpha\beta\gamma\delta}
	N_{\gamma\delta}.
\end{equation}
Here 
\begin{equation}
J^{(a)(b)} = m^{(a)(b)}_A j_A,
\end{equation}
and
\begin{equation}
H^{(a)(b)(c)(d)} = \frac{1}{2}\, m^{(a)(b)}_A\,  M_{AB}\, 
m^{*(c)(d)}_B,
\end{equation} 
where $M_{AB}$ is the matrix of absorption and rotation coefficients that appears in (\ref{poltrans}).
For models in which absorption and rotation can be described in terms of the classical response of the plasma, the tensor $H^{(a)(b)(c)(d)}$ is directly related to the components of the plasma dielectric tensor; the relationship is given in \citet{gammie:2012} (their eq. 64).  This form of the polarized transport equation is equivalent to that used in \citet{bb:2004}.  

\section{Numerical Methods}\label{sect:numerical}

Equation (\ref{eq:covtrans2}) might seem an unpromising start for a numerical integration scheme, since the basic equation is complicated and one has to integrate the 16 real degrees of freedom in $N^{\alpha\beta}$ compared to the 4 real degrees of freedom in a Stokes basis representation of the radiation field.  Still, $N^{\alpha\beta}$ is manifestly covariant and conceptually simple: the tensor notation takes care of all frame transformations automatically.  Also, the integration of additional degrees of freedom is, it turns out, not the leading cost in polarized ray-tracing calculations.

Our second-order integration strategy splits (\ref{eq:covtrans2}) into two parts.  The first part incorporates parallel transport: it uses the LHS and the {\em first} two terms on the RHS to parallel transport the polarized radiation field in the coordinate basis.  The second part incorporates emission, absorption, and Faraday rotation: it transforms the LHS and the {\em second} two terms on the RHS into the Stokes basis in the frame of the plasma, where the transfer coefficients are most naturally evaluated.   These latter terms yield
\begin{equation}\label{spoltrans}
\frac{d}{d\lambda}
\begin{pmatrix} I \\ Q \\ U \\ V \end{pmatrix}
 = \begin{pmatrix} j_{I} \\ j_{Q} \\ j_{U} \\ j_{V} \end{pmatrix}
- \begin{pmatrix} 
\alpha_{I} & \alpha_{Q} & \alpha_{U} & \alpha_{V} \\
\alpha_{Q} & \alpha_{I} & \rho_{V} & -\rho_{U} \\
\alpha_{U} & -\rho_{V} & \alpha_{I} & \rho_{Q} \\
\alpha_{V} & \rho_{U} & -\rho_{Q} & \alpha_{I}  
\end{pmatrix}
\begin{pmatrix} I \\ Q \\ U \\ V \end{pmatrix} + \ldots
\end{equation}
where, again, the absence of subscript $\nu$ implies that a term appears in invariant form, i.e. $\rho_V = \nu \rho_{\nu,V}$ and the derivative is understood to follow an individual photon in frequency space. 

What technique should one use to evolve (\ref{spoltrans})?  One consideration
is computational expense when the Faraday or absorption depth is large.  Most
explicit schemes will be limited by $\Delta\lambda \lesssim {\rm MIN}(
1/\alpha_A, 1/\rho_A)$.  Many $\lambda$-steps are then required to cross the
system, even if the transfer coefficients change smoothly.  For example, the
Faraday rotation in some models of Sgr A* and M87 at $1.3$mm is very large
(e.g., \citealt{moscibrodzka:2017}), so a simple second-order integration
scheme would require many $\lambda$-steps to cross the system as it would be
limited to rotating the electric vector polarization angle (EVPA) by $O(1)$
radian per step.  A second consideration is that the source models that
motivated the development of {\tt ipole} are derived from numerical
simulations, which have an irreducible granularity because they represent the
physical variables on a grid.  It makes no sense to meticulously integrate
(\ref{spoltrans}) across a single simulation zone when the structure of 
the model inside
the zone is known only up to truncation error.  Still, even in this case a stable and physically sensible evolution of (\ref{spoltrans}) is desirable.

It would therefore be helpful to use a numerical technique that takes
advantage of analytic solutions to (\ref{spoltrans}) assuming constant
transfer coefficients.  Indeed, this is what the DELO family of polarized
transfer solvers does (\citealt{rees:1989}, \citealt{janett:2017}) while making particular assumptions about conditions in the source. More generally, \citet{LDI:1985} (hereafter \LDI) found an elegant, formal solution of
the problem expressed in terms of an integral along the line of sight.  This
solution can be also found in \citet{peraiah:2001} (notice that section 12.6
contains a few typographical errors in their equations: 12.6.10, 
12.6.27, 12.6.29, 12.6.31, 12.6.32) and, partially, in \citet{dexter:2016}
(contains a typographical error in [D5], $M_3$[0,2] should be $\Lambda_1 \alpha_U + \sigma \Lambda_2 \rho_U$).  Our integration scheme uses the \LDI\, solution in explicit form.

The explicit general polarized transport solution with constant coefficients 
can be obtained following \LDI, who write the transfer equation (\ref{spoltrans})
in the form
\begin{equation}
\frac{d S_A}{d\lambda} = j_A - K_{AB} S_B,
\end{equation}
where we have recast the equation using our index notation, substituted $j_A$ for their ${\bf K S}$ (${\bf S}$ is \LDI's source function vector), and cast the basic equation in invariant form with independent variable $\lambda$ rather than $s$.  The formal solution is
\begin{equation}
S_A(\lambda) = \int^\lambda_{\lambda_0} O_{AB}(\lambda-\lambda') j_B d\lambda' + O_{AB}(\lambda-\lambda_0) S_B(\lambda_0),
\end{equation}
where $O_{AB}$ is given by their eq. (10).\footnote{There is a 
typographical error in $M_4[1,1]$; $n_Q$ should
read $\eta_Q$.}  This formal solution still requires evaluation of the
integral to put it in a form suitable for numerical integration.  Defining
\begin{equation}
P_{AB} \equiv \int^\lambda O_{AB},
\end{equation}
the formal solution for constant coefficients is
\begin{equation}\label{eq:fullsol}
S_A(\lambda) = P_{AB}(\lambda-\lambda_0) j_B + O_{AB}(\lambda-\lambda_0) S_B(\lambda_0).
\end{equation}
Integrating \LDI\, eq.(10), one finds
\begin{multline}
P_{AB} = -\Lambda_1 f_1 M_{3,AB} + \frac{\alpha_I f_1}{2}   (M_{1,AB} + M_{4,AB}) \\
	    \qquad  + \Lambda_2 f_2 M_{2,AB} + \frac{\alpha_I f_2}{2}   (M_{1,AB} - M_{4,AB}) \\
        - e^{-\alpha_I \Delta \lambda}  \,\, \times \\
        \Bigg\{\Big[-\Lambda_1 f_1 M_{3,AB} + \frac{\alpha_I f_1}{2}   (M_{1,AB} + M_{4,AB})\Big] \cosh(\Lambda_1 \Delta \lambda) \\
        + \Big[-\Lambda_2 f_2 M_{2,AB} + \frac{\alpha_I f_2}{2}   (M_{1,AB} - M_{4,AB})\Big] \cos(\Lambda_2 \Delta \lambda) \\
        + \Big[-\alpha_I f_2 M_{2,AB} - \frac{\Lambda_2 f_2}{2}  (M_{1,AB} - M_{4,AB})\Big] \sin(\Lambda_2 \Delta \lambda) \\
        - \Big[\alpha_1 f_1 M_{3,AB} - \frac{\Lambda_1 f_1}{2}  (M_{1,AB} + M_{4,AB})\Big] \sinh(\Lambda_1 \Delta \lambda) \Bigg\}.
\end{multline}
Here $\Delta \lambda \equiv \lambda - \lambda_0$ and the notation follows \LDI\, including the definition of the $4 \times 4$ matrices $M$, except that we have introduced $f_1 \equiv (\alpha_I^2 - \Lambda_1^2)^{-1}$ and $f_2 \equiv (\alpha_I^2 + \Lambda_2^2)^{-1}$, and our $\alpha_S$ is their $\eta_S$. 
The reader is referred to \LDI,
or the publicly released code, for a complete account of the solution.   

Solution (\ref{eq:fullsol}) is complicated and difficult to manipulate algebraically.  For convenience, we provide two  special solutions in the appendix, for when only Faraday conversion is present and for when only absorption and emission are present.

\subsection{Integration scheme}

The full image-generation routine proceeds as follows.  The basic notion is identical to the publicly available {\tt ibothros} code. \footnote{\tt https://github.com/AFD-Illinois/ibothros2d}  An observer is placed at a fixed spacetime event and given a four-velocity and a ``camera'' which is defined via an orthonormal tetrad at the observer.  The camera has pixels, which form a regular grid in angle.  If the camera is pointed at the black hole, the central point of the frame is defined so that photons arriving at that point have zero angular momentum.  Geodesics are integrated backwards from the center of each pixel through the source until a stopping condition is met (the stopping condition is problem dependent).  The coordinates and wavevectors along the geodesic are recorded during the backwards integration.  

The transfer equation is then integrated {\em forward} along the geodesic to the camera.  Begin by setting the Stokes vector using a boundary condition, usually $S_A = 0$.  Then convert $S_A$ to $N^{\alpha\beta}$ using \ref{eq:StoN} and evolve $N^{\alpha\beta}$ forward along the geodesic.

(1) Evaluate the connection coefficient at the initial position and parallel transport $N^{\alpha\beta}$ by a half step using the first two terms in (\ref{eq:covtrans2}).  This is done using a simple second-order integrator.  Since the rest of the scheme is second order there is no point in going to higher order. 

(2) Erect an orthonormal tetrad $e^\mu_{(a)}$ in the plasma frame at the half-step position, with $e^\mu_{(0)} = u^\mu$, the plasma four-velocity, $e^\mu_{(3)}$ parallel to the spatial component of the wavevector in the plasma frame, and $e^\mu_{(1)}$ and $e^\mu_{(2)}$ in the plane perpendicular to both.  In most problems of interest to us synchrotron emission is important, so ordinarily we require that $e_{(2)}$ is in the plane formed by the wavevector and the magnetic field in the plasma frame.  Adopt the convention that $Q > 0$ corresponds to linear polarization in the $e{(1)}$ direction.  Then for synchrotron emission and absorption, $j_U = \alpha_U = 0$, and if Faraday conversion is due to a magnetized plasma then $\rho_U = 0$.   

(3) Evaluate the transfer coefficients in the tetrad frame.

(4) Project $N^{\alpha\beta}$ into $S_A$ in the tetrad frame.

(5) Evolve the Stokes vector by a full $\lambda$ step using the analytic solution (\ref{eq:fullsol}).

(6) Transform $S_A$ back into $N^{\alpha\beta}$ using the tetrad basis.

(7) Parallel transport $N^{\alpha\beta}$ by another half-step.

Substeps (7) and (1) can be combined without formal loss of accuracy if a half-step is taken at the beginning and end of the integration and the stepsize is constant.  The initial and final half-step can also be dropped without loss of accuracy if they occur in regions where there is no substantial evolution of $N^{\alpha\beta}$.

Finally, the Stokes parameters are observed in the camera tetrad using equation (\ref{eq:NtoS}) and recorded at each pixel.

\section{Tests of numerical scheme}\label{sect:test}

\subsection{Tests of transport step in non-trivial geometries}

\begin{figure}
\begin{center}
\includegraphics[width=0.5\textwidth,angle=0]{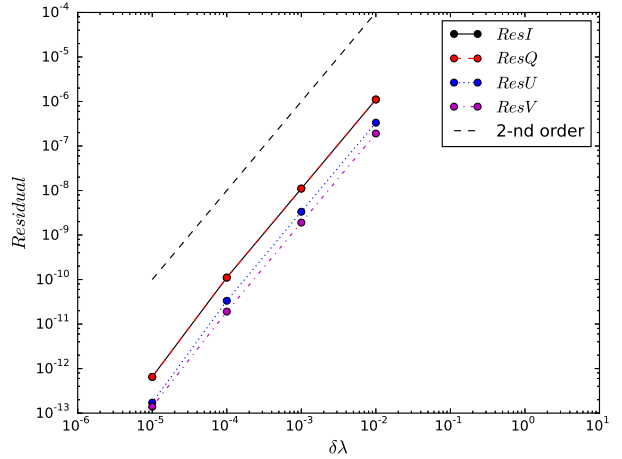}
\caption{Transport-step test no 1: 
Convergence of the transport-step in vacuum when polarized light is
transported in Minkowski space with
``snake'' Cartesian coordinates. The residuals between the Stokes 
parameters at the beginning and at the end of integration path are shown as a
function of the step-size. The transport scheme converges at second order.}
\label{fig:test1}
\end{center}
\end{figure}

\begin{figure}
\begin{center}
\includegraphics[width=0.5\textwidth,angle=0]{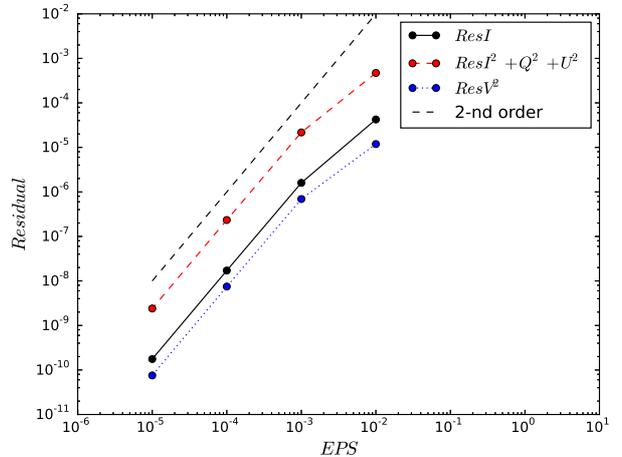}
\caption{Transport-step test no 2: Convergence of the transport-step in vacuum
  in a near vicinity of the event horizon of the Kerr black hole. Here we show
  residuals of invariant quantities between initial and final integration
  point as a function of parameter describing the step-size. The transport scheme converges at second order.}
\label{fig:test2}
\end{center}
\end{figure}

The parallel transport of $k^\mu$ and $N^{\alpha\beta}$ is realized using a second-order integrator  (meaning the single-step error is $O(\Delta\lambda^3)$ and therefore the error at the camera is $O(\Delta\lambda^2)$ after integrating over $O(\Delta\lambda^{-1})$ steps). Parallel transport tests considered in this section assume a non-zero initial $N^{\alpha\beta}$ and transport $N^{\alpha\beta}$ in vacuum (i.e. we are solving (\ref{eq:covtrans2}) assuming that all transport coefficients vanish). We test the transport of polarized light in i) Minkowski spacetime using snake Cartesian coordinates (see section 4.3 in \citealt{white:2016}) and ii) Kerr spacetime described by modified Kerr-Schild coordinates \citep{gammie:2003}.

i) The snake coordinates ($X^0,X^1,X^2,X^3$) vary periodically with Minkowski position ($t,x,y,z$). The two coordinate systems are related via $(X^0,X^1,X^2,X^3)=(t,x,y+a \sin(k x),z)$, where $a=0.3$ and $k=\pi/2$ are default parameters. For $\delta=a k\sin(k X^1)$ the geometry is described by the following metric tensor:
\begin{equation}
g_{\mu \nu}=
 \begin{pmatrix}
-1 & 0 & 0 & 0 \\
0 & \sqrt{1+\delta^2} & -\delta & 0 \\
0 & -\delta & 1 & 0 \\
0 & 0 & 0 & 1
\end{pmatrix}.
\end{equation}
In snake coordinates (\ref{eq:covtrans2}) has source terms because the connection coefficient $\Gamma^{2}_{11}$ does not vanish.  We find $\Gamma^{2}_{11}$ by numerically differentiating the metric tensor (a facility for obtaining the connection coefficients by numerical differentiation of the metric is provided in the default, public version of the code).

In flat spacetime, in the absence of emitting and absorbing matter, the Stokes
parameters should remain constant when measured in a parallel transported tetrad
attached to $k^\mu$ (here $Q$ and $U$ are read out in a tetrad in which the basis vectors perpendicular to $k^\mu$ are aligned with the snake coordinates). Figure~\ref{fig:test1} displays residuals of Stokes
parameters extracted from $N^{\alpha\beta}$ at $x_{\rm final}=3$ (where
$x_{\rm final}$ is the end of the integration path that starts at $x_{\rm
  init}=0$) with respect to their initial values as a function of the constant
step size. As expected, the residuals decrease as $(\Delta\lambda)^2$. 

\begin{figure*}
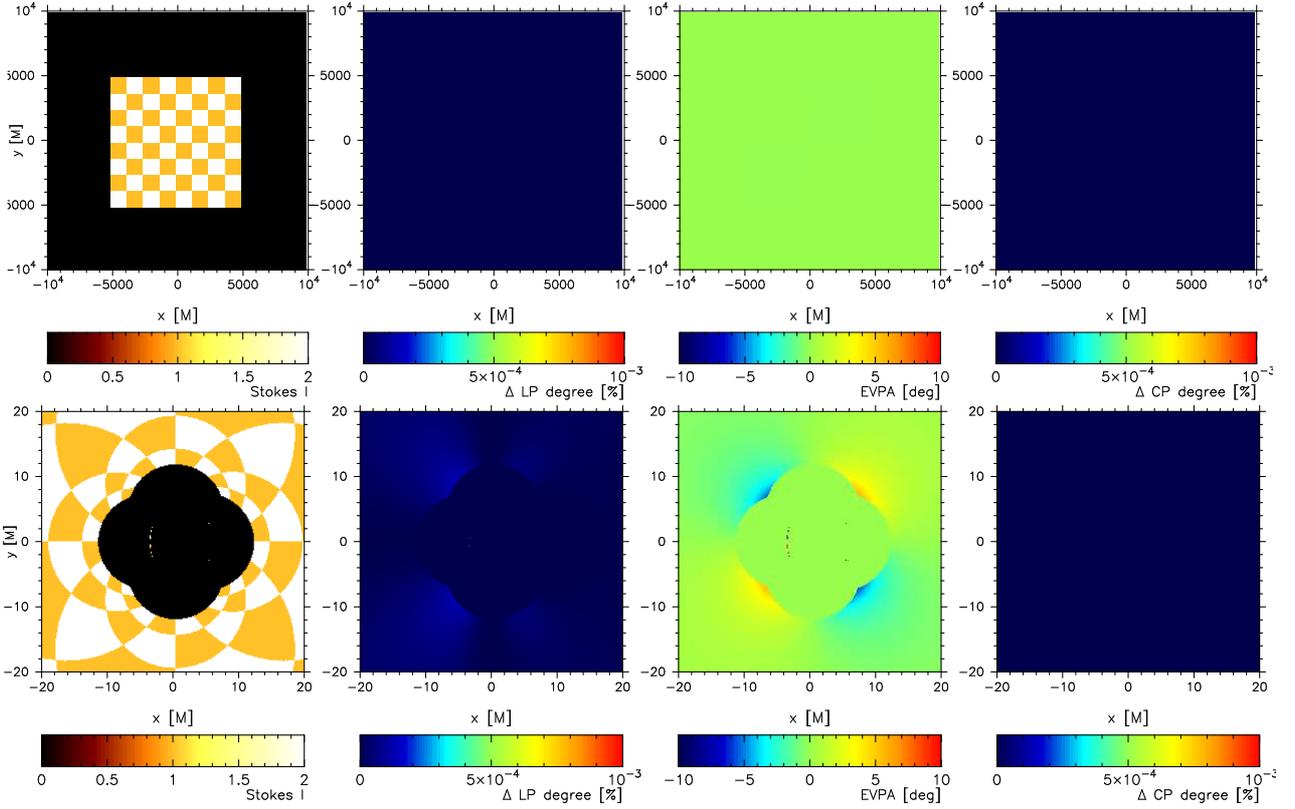

\begin{center}
\includegraphics[width=0.3\textwidth,angle=-90]{fig3a.eps}\\
\includegraphics[width=0.3\textwidth,angle=-90]{fig3b.eps}
\caption{Transport-step test no 3: image of uniformly polarized screen (of size equal $10^4\times10^4$ M) behind the spinning black hole. Observer's viewing angle is 90 degrees with respect to the black hole spin axis. The upper panels show the image of the screen for a large field-of-view
  to show the problem setup. At these scales the image is barely affected by the gravitational lensing and EVPA is zero. The lower panels show the zoom-in of the upper panels onto inner regions where lensing is significant and therefore distorts the checkerboard pattern. Panels from left to right show:
Stokes I (transport-step invariant), the change of linear polarization degree (transport-step invariant), EVPA and the change of circular polarization degree (which square value is also the transport-step invariant). Here we see some rotation of polarization angle.}\label{fig:test34}
\end{center}
\end{figure*}

ii) In the second test we check performance of the parallel transport in Kerr
metric in modified Kerr-Schild coordinates. Here the integration is carried
out along geodesics that pass the black hole event horizon with an impact
parameter of 5 $GM/c^2$.  The black hole dimensionless angular momentum
parameter $a/M= 0.9375$.  First, we checked that during integration the
parallel transported rank-2 tensor $N^{\alpha\beta}$ remains
Hermitian. Second, we check three invariant quantities along the ray: Stokes $I$,
combination of Stokes parameters $I^2+Q^2+U^2$, and $V^2$.
Figure~\ref{fig:test2} shows residuals between initial invariant quantities
and the ones measured at the end of geodesics integration (at large distance
from the black hole). The residuals are shown as a function of step-size
control parameter ${\tt EPS}$. The residuals evidently decrease as ${\tt EPS}^2$. Our
second order scheme has single precision accuracy for ${\tt EPS} \lesssim 10^{-3}$,
which is the value we typically use in {\tt ibothros} when generating mock
observations of a GRMHD simulation.

Figure~\ref{fig:test2} demonstrates the convergence of the transport scheme in
Kerr metric along a single geodesic. We are interested in constructing images
of an accreting black hole at a camera located far from the hole. In
the third test we demonstrate the accuracy of the transport step when
constructing such images. The observer is located at $r_{\rm cam}=10^6 \rg$.
The observer's line of sight is oriented at 90 degrees with respect to the black hole spin axis. We set a screen producing a uniformly polarized radiation at $r=10^4 \rg$
behind the black hole. The screen has size $10^4 \times 10^4 \rg$.
The Stokes parameters at the screen are generated using the same tetrad construction procedure as that used for the camera.
A checkerboard pattern in Stokes I is introduced to help visualize how gravitational lensing distorts the background screen. The degree of linear polarization $LP=\sqrt{Q^2+U^2}/I$=100 per cent and degree of circular polarization $CP=|V|/I$=25 per cent are constant across the entire screen.

Figure~\ref{fig:test34} shows how a Kerr black hole distorts the background checkerboard
pattern. Top and bottom panels show the same model at large and small scales, respectively. 
For a large field-of-view the pattern is only weakly affected by the gravitational field of the black hole.
For a smaller field-of-view the pattern is strongly lenses and the image of the screen edges resemble a four-leaf clover. In vacuum Stokes I, $I^2+Q^2+U^2$ and $V^2$ are invariant, and
consequently the linear and circular polarization fractions are invariant. 
We find that these
radiative transport invariants are conserved for any given ray that reaches
the observer with accuracy better than 0.01 percent. 
Notice however that the
polarization angle $EVPA$ is a function of ray impact
parameter. The $EVPA$ rotation is expected because of gravitational Faraday
rotation (e.g., \citealt{ishihara:1988}, \citealt{sereno:2005}).

\subsection{Tests source step combined with transport step}

Next we test the part of the code that evolves the Stokes parameters.
\citet{dexter:2016} (Appendix C) presents two cases where (\ref{poltrans}) has
an analytic solution in a simple functional form. These two examples are in
Minkowski spacetime and either $j_{IQ}\neq 0$ and $\alpha_{IQ}\neq 0$ or
$j_{QUV}\neq 0$ and $\rho_{QV}\neq 0$. Other transfer coefficients are set to
zero.  Here we repeat these two tests in the snake coordinates.

\begin{figure*}
\begin{center}
\includegraphics[width=0.48\textwidth,angle=0]{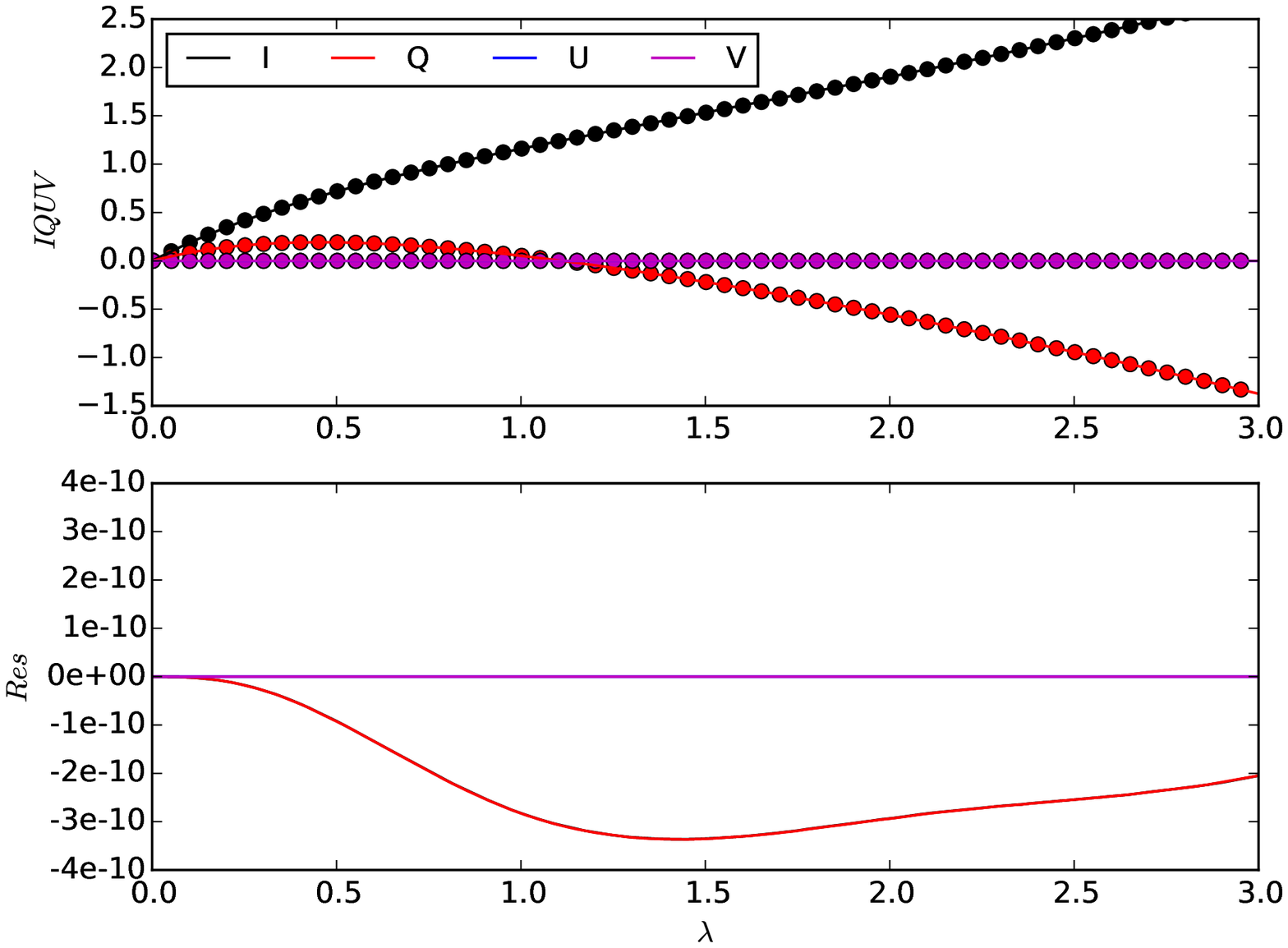}
\includegraphics[width=0.48\textwidth,angle=0]{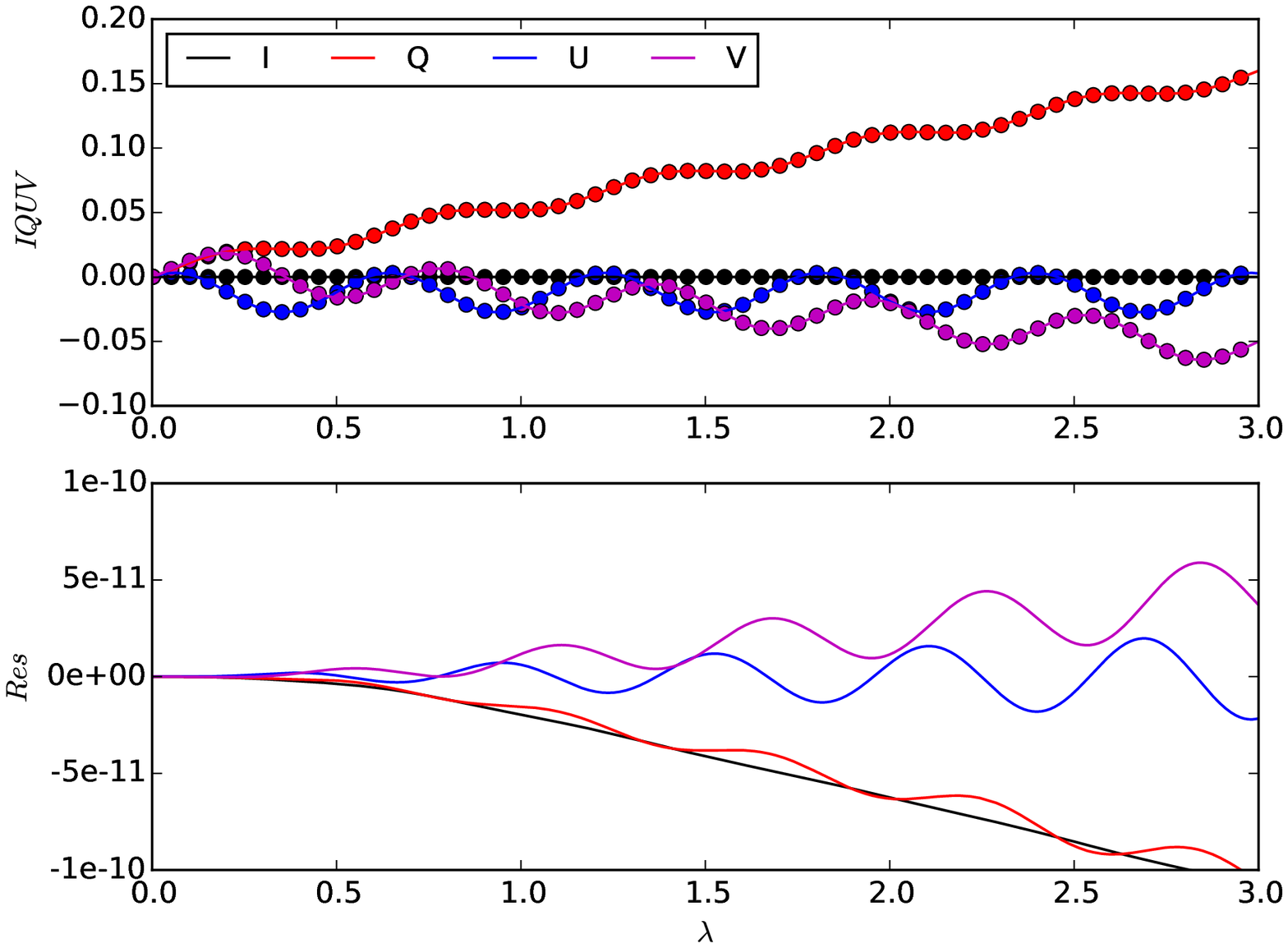}
\caption{Source-step tests no 1 and 2: Numerical (points) and analytic (lines)
  solutions of radiative transport of polarized light in 1D in snake
  coordinates. Left panel: test of emission and absorption of stokes $I$ and
  $Q$. Right panel: test of emission and rotation of Stokes $Q$, $U$, and $V$
  (see text for details).}
\label{fig:source12}
\end{center}
\end{figure*}

In the first test, $j_{IQ}=(2,1)$ and $\alpha_{IQ}=(1,1.2)$ are the only
non-zero elements on the RHS of (\ref{eq:covtrans2}) (apart from the
$\Gamma^2_{11}$ coefficient needed for parallel transport in snake
coordinates). 
Figure~\ref{fig:source12} (left panel) compares the {\tt ipole}
numerical and known analytic solutions. For step size
$\Delta\lambda=10^{-3}$ (although for constant transfer coefficients our errors do not depend on the step size) the residuals between numerical and analytic model are
better than single-precision accuracy. 

In the second test, $j_{QUV}=(0.1,0.1,0.1)$ and
$\rho_{QV}=(10,-4)$. Figure~\ref{fig:source12} (right panel) shows the
results. Here, the residuals between numerical and analytic solution are even
smaller compared to the emission/absorption test in the left panel. 
The errors oscillate and grow with $\lambda$.

\subsection{Comparison of {\tt ipole} and {\tt grtrans}}

\subsubsection{Relativistic plasma in Minkowski space}

Next, we consider a radiative transfer problem in a slab of relativistically
hot, magnetized plasma with varying plasma density, temperature, magnetic
field strength and magnetic field direction. The plasma is emitting,
absorbing, and Faraday rotating/converting polarized synchrotron radiation.
This problem has no analytic solution, so we test by comparison with ${\tt
  grtrans}$.  

We use the same $j_S$, $\alpha_S$ and $\rho_S$ as those in ${\tt grtrans}$. 
 The exact formulae for emissivity, absorptivity, and
  rotativity are written down in \citet{dexter:2016} in appendices A1 and B2. 
The expressions for Faraday rotativities follow \citet{shcherbakov:2008}. 
Each coefficient is a distinct function of plasma density,
temperature, magnetic field strength, photon frequency, and orientation of the
magnetic field with respect to $k^\mu$. This test also allows us to test our
implementations of units, as both codes produce results in cgs units.

\begin{figure*}
\begin{center}
\includegraphics[width=0.9\textwidth,angle=0]{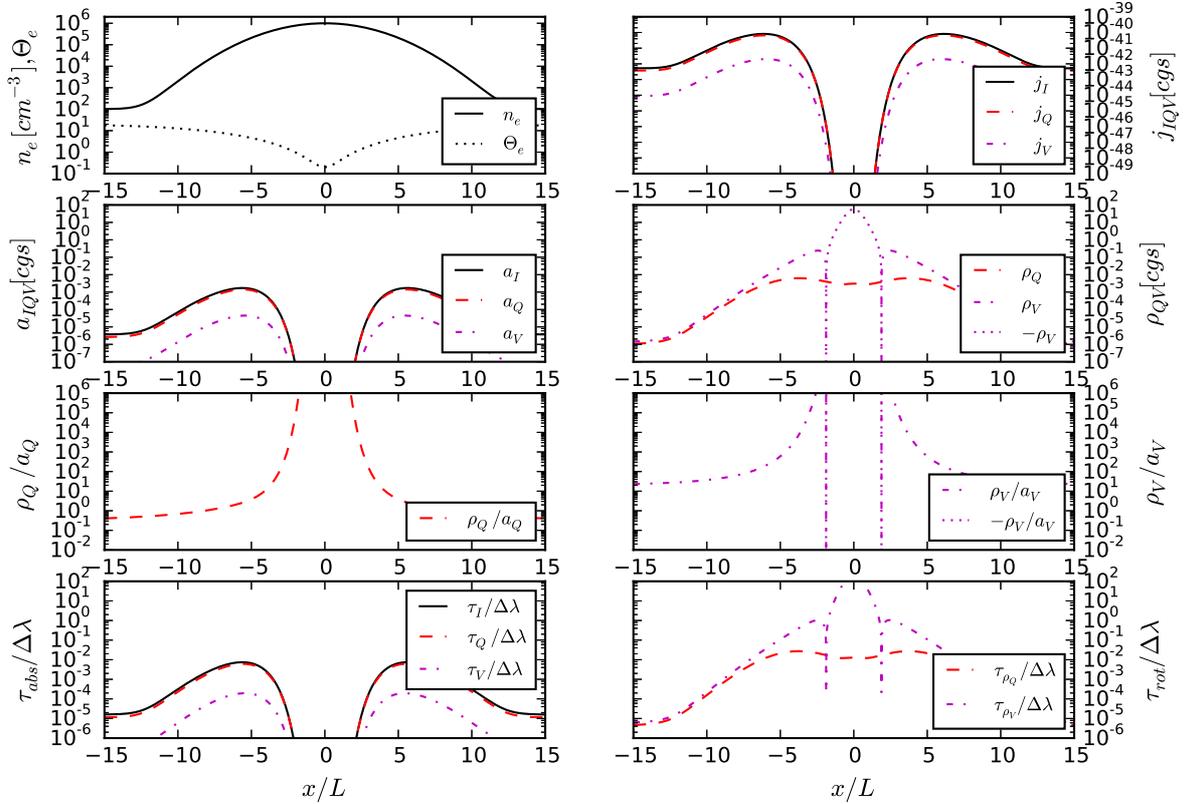}
\caption{Source-step test no 3: structure of the slab of plasma together with
  all synchrotron emissivities, absorptivities and rotativities, their ratios
  and optical and Faraday thickness per one step along the integration path.
}\label{fig:prof}
\end{center}
\end{figure*}

We integrate (\ref{eq:covtrans2}) along the $x$-axis from $x=-15 L$ to $x=15
L$, where $L=10^{15}$ cm.  The plasma electron number density varies smoothly with $x$ as
\begin{equation}
n_e=n_0\left[1+A\, \exp^{-(x/L)^2/\sigma_x^2}\right],
\end{equation} 
where $n_0=10^2$, $A=10^4$, and $\sigma_x=4$ are free parameters. The
electrons have a relativistic, thermal (Maxwell-J{\"u}ttner) distribution
function described by dimensionless electron temperature
$\Theta_e=k_BT_e/(m_e c^2)$. Electron temperature is also a smooth, slowly
changing function of x:
\begin{equation}
\Theta_e=\Theta_{e,0}\left[1+A \,\exp^{-(x/L)^2/\sigma_x^2}\right],
\end{equation}
where $\Theta_{e,0}=20$, $A=-0.99$, and $\sigma_x=10$ are free parameters. The density and temperature profiles are shown
in Figure~\ref{fig:prof} (top left panel). For simplicity, we assume that
magnetic field strength B=30 Gauss and its orientation $\theta=60$ degrees are
constant along the integration path.  Also the spatial components of the
plasma four-velocity are zero. The radiative transfer equations are integrated
for a photon with frequency of 230 GHz.  The invariant synchrotron
emissivities, absorptivities, and rotativities and their ratios along the
integration path are shown in Figure~\ref{fig:prof}.  Two bottom panels in
Figure~\ref{fig:prof} show the optical and Faraday optical thickness per
integration step.

\begin{figure*}
\begin{center}
\includegraphics[width=0.9\textwidth,angle=0]{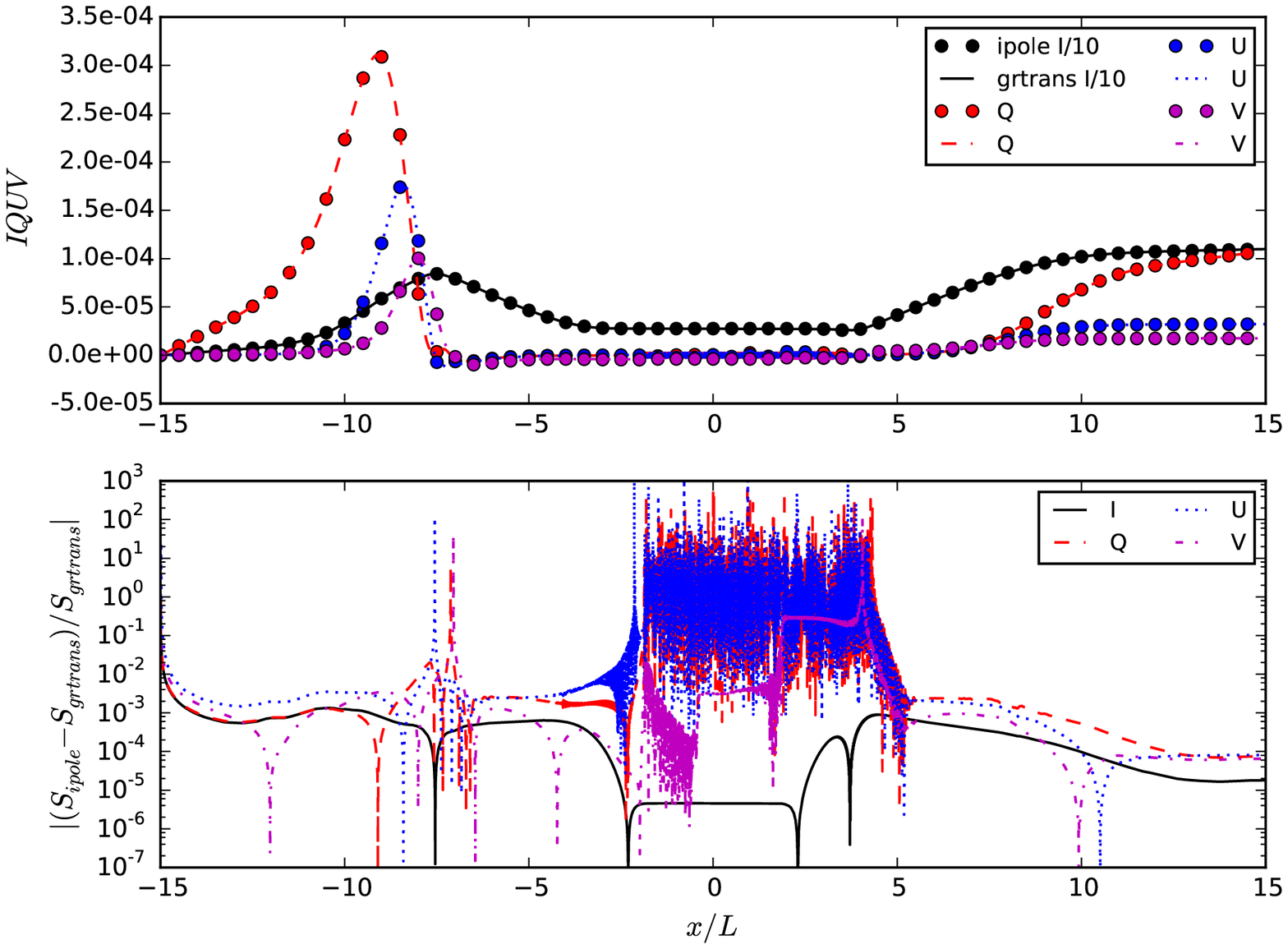}
\caption{Source-step test no 3: Comparison of radiative transfer integration
  with {\tt ipole} and {\tt grtrans} schemes. The largest difference is in
  Stokes Q, U, and V in regions of large Faraday optical depth, $\tau_F \gg
  1$.  Otherwise the relative difference between the codes is less than 1\%.
}\label{fig:phys}
\end{center}
\end{figure*}

Figure~\ref{fig:phys} shows radiative transfer solutions through the plasma
shown in Figure~\ref{fig:prof}. Here all Stokes parameters are shown in cgs
units as produced by {\tt ipole} and {\tt grtrans}.  The codes agree with each
other well, except for Stokes Q and U in regions with high Faraday depth (between
$x=-3$ L and $x=5$ L) where $Q,U$ and $V$ are small.

\subsubsection{Polarized transport in hot accretion flows onto a black hole}

\begin{figure*}
\begin{center}
\includegraphics[width=0.6\textwidth,angle=-90]{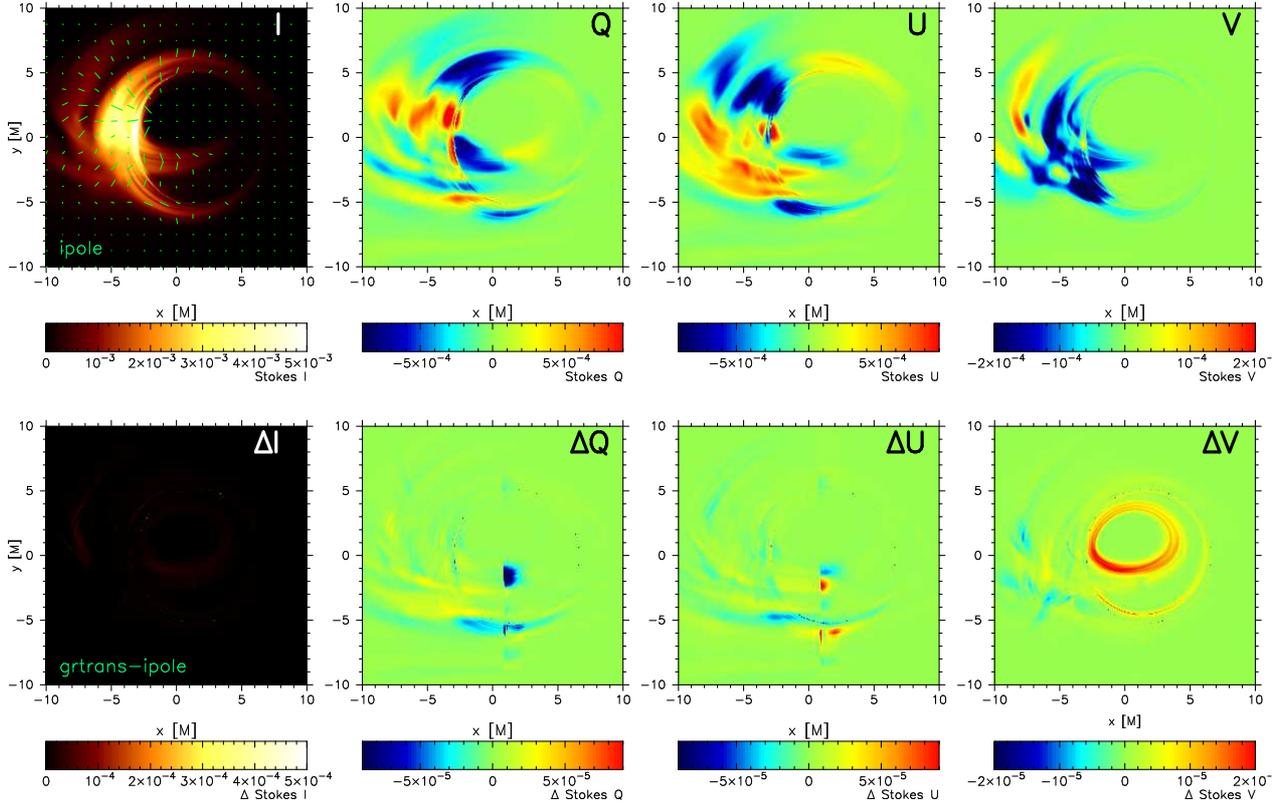}
\caption{Polarized millimetre images of a Kerr black hole ($a/M = 0.9375$) accreting
  matter. The dynamics of magnetized plasma around the black hole is a 3D
  GRMHD model of thick accretion disk with turbulent magnetic fields. Top panels show the Stokes I
  (with green ticks indicating the direction of EVPA and the length of each tick being
  proportional to a local $\sqrt{Q^2+U^2}$), Stokes Q, U and V. The dark circular
  shadow in the Stokes I map is the shadow of the black hole event
  horizon. Bottom panels show corresponding residuals between Stokes parameters in
  {\tt ipole} and {\tt grtrans} output. The field of view in all images is
  $20\times20 \,\rg$ with resolution of $256 \times 256$ pixels and the
  observer's line of sight is 60 degrees away from the black hole spin axis.}\label{fig:grmhd}
\end{center}
\end{figure*}

In Figure~\ref{fig:grmhd} (upper panels) we present an example of {\tt ipole} 
polarized images of hot, magnetized turbulent accretion flow around a Kerr black hole.  
The underlying plasma accretion flow model is a 3D GRMHD Fishbone-Moncrief torus simulation 
carried out with {\tt harm3d}
code (\citealt{gammie:2003}, \citealt{noble:2006}). The simulation data is
converted from the code units to cgs units assuming black hole mass
$M_{\rm BH}=6.2 \times 10^9 \, {\rm M_\odot}$ and the mass accretion
rate onto the black hole $\dot{M} = 1.1 \times 10^{-4} {\rm M_\odot yr^{-1}}$.
The model requires a prescription for
electron temperature; we assume that electron temperature equals
proton temperatures in the entire computational domain. 
In this test, the observer is located at distance of r=1000 M from the black hole and the line of sight
  is at 60 degrees to the black hole spin axis.

We repeat the radiative transport calculation through the same simulation
snapshot using {\tt grtrans}.  Figure~\ref{fig:grmhd} (lower panels) shows
difference between {\tt ipole} and {\tt grtrans} outputs. The differences are
small.  We quantify the difference between images using mean square error
defined as $MSE_S=\sum_{ij} (S_{\rm ipole}-S_{\rm grtrans})^2/\sum_{ij} S_{\rm                                            
  grtrans}^2$, where S is the Stokes parameter and summations are done over
all image pixels. The results
are $MSE_{\rm I}=4.38 \times 10^{-5}$, $MSE_{\rm Q}=1.44 \times 10^{-3}$, $MSE_{\rm U}=9                                                    
\times10^{-4}$, and $MSE_{\rm V}=3.92 \times 10^{-3}$ for stepping parameter ${\tt EPS}=0.0025$.
One can also quantify the agreement between two corresponding Stokes maps
using the image quality index $Q_{\rm idx}$ \citep{wang:2002}.
We find $Q_{\rm idx}(I,Q,U,V)$=(0.999968,0.999173,0.998880,0.995589), where 
 $Q_{\rm idx}$=1 would mean that two images are identical, which confirms strong consistency between corresponding Stokes maps.
We conclude that the agreement between the two codes
is excellent even for a very complex problems.

In the future we will test the convergence of radiative transfer simulations
through various GRMHD simulations as a function of the step size along
geodesics and as a function of number of pixels in the images. In our example calculation
we also assumed that the dynamical simulations are static and the plasma conditions do not change 
as the light propagates through it (the ``fast light'' approximation). Near a black
hole event horizon, however, the light crossing time is comparable to the
dynamical time.  It is important to quantify how sensitive the
observed Stokes parameters are to spatial and temporal resolution (i.e., cadences of data dumps) of the
numerical simulations, but such a study is beyond the scope of the present paper.

\section{Summary}\label{sect:summary}

We have designed a numerical scheme capable of integrating relativistic
polarized radiative transfer equations by ray tracing in non-trivial spacetimes and in optical and Faraday thick plasmas. We have demonstrated that the integration scheme is stable and accurate and can reproduce known analytic solutions. The code has been tested on scaled problems and on dimensional problems to test the unit system. Our results agree with results from J. Dexter's independent code, {\tt grtrans}.

We plan to extend {\tt ipole} to include scattering within a Monte Carlo
framework, so that it can make predictions for a broader range of sources and
photon energies (\citealt{connors:1977}, \citealt{connors:1980}), 
motivated by results from INTEGRAL and the future X-ray polarization mission IXPE.

Does the code run efficiently, i.e. how fast is it?  Our reference machine is a two socket Intel Xeon E5-2660 at 2.6 GHz, which has a total of 20 physical cores (40 with hyperthreading).  We compile the code with a version of {\tt h5cc} that uses {\tt gcc 4.8.5} and {\tt -Ofast --fopenmp}.  We find that a $256^2$ fully polarized image with a $40 GM/c^2$ field of view, using data from a {\tt harm3d} model, completes in $8.6s$ clocktime and $253s$ cpu time.  On a single core, the code completes in $136s$, for an average speed of $480$ rays per second.  This speed is similar to that of {\tt ibothros2d}.

{\tt ipole} is publicly available at {\tt https://github.com/moscibrodzka/ipole} (note: the code will be released simultaneously with publication).
 
\section*{Acknowledgements}

M. Mo\'scibrodzka acknowledges support from the ERC Synergy Grant
``BlackHoleCam-Imaging the Event Horizon of Black Holes'' (Grant
610058). C. F. Gammie acknowledges support from NSF grant AST-1333612 and 
AST-1716327, a Romano Professorial Scholarship, and the hospitality of the Flatiron 
Institute's Center for Computational Astrophysics where some of this work was 
completed.  The authors thank J. Dexter for his extensive help and guidance.
The authors also thank B. Ryan, G. Janett, and the referee for their comments.





\begin{thebibliography}{}
\makeatletter
\relax
\def\mn@urlcharsother{\let\do\@makeother \do\$\do\&\do\#\do\^\do\_\do\%\do\~}
\def\mn@doi{\begingroup\mn@urlcharsother \@ifnextchar [ {\mn@doi@}
  {\mn@doi@[]}}
\def\mn@doi@[#1]#2{\def\@tempa{#1}\ifx\@tempa\@empty \href
  {http://dx.doi.org/#2} {doi:#2}\else \href {http://dx.doi.org/#2} {#1}\fi
  \endgroup}
\def\mn@eprint#1#2{\mn@eprint@#1:#2::\@nil}
\def\mn@eprint@arXiv#1{\href {http://arxiv.org/abs/#1} {{\tt arXiv:#1}}}
\def\mn@eprint@dblp#1{\href {http://dblp.uni-trier.de/rec/bibtex/#1.xml}
  {dblp:#1}}
\def\mn@eprint@#1:#2:#3:#4\@nil{\def\@tempa {#1}\def\@tempb {#2}\def\@tempc
  {#3}\ifx \@tempc \@empty \let \@tempc \@tempb \let \@tempb \@tempa \fi \ifx
  \@tempb \@empty \def\@tempb {arXiv}\fi \@ifundefined
  {mn@eprint@\@tempb}{\@tempb:\@tempc}{\expandafter \expandafter \csname
  mn@eprint@\@tempb\endcsname \expandafter{\@tempc}}}

\bibitem[\protect\citeauthoryear{{Balazs}}{{Balazs}}{1958}]{balazs:1958}
{Balazs} N.~L.,  1958, \mn@doi [\apj] {10.1086/146553}, \href
  {http://esoads.eso.org/abs/1958ApJ...128..398B} {128, 398}

\bibitem[\protect\citeauthoryear{{Ball}, {{\"O}zel}, {Psaltis}  \&
  {Chan}}{{Ball} et~al.}{2016}]{ball:2016}
{Ball} D.,  {{\"O}zel} F.,  {Psaltis} D.,   {Chan} C.-k.,  2016, \mn@doi [\apj]
  {10.3847/0004-637X/826/1/77}, \href
  {http://esoads.eso.org/abs/2016ApJ...826...77B} {826, 77}

\bibitem[\protect\citeauthoryear{{Bower}, {Wright}, {Falcke}  \&
  {Backer}}{{Bower} et~al.}{2003}]{bower:2003}
{Bower} G.~C.,  {Wright} M.~C.~H.,  {Falcke} H.,   {Backer} D.~C.,  2003,
  \mn@doi [\apj] {10.1086/373989}, \href
  {http://esoads.eso.org/abs/2003ApJ...588..331B} {588, 331}

\bibitem[\protect\citeauthoryear{{Bower}, {Falcke}, {Wright}  \&
  {Backer}}{{Bower} et~al.}{2005}]{bower:2005}
{Bower} G.~C.,  {Falcke} H.,  {Wright} M.~C.,   {Backer} D.~C.,  2005, \mn@doi
  [\apjl] {10.1086/427498}, \href
  {http://esoads.eso.org/abs/2005ApJ...618L..29B} {618, L29}

\bibitem[\protect\citeauthoryear{{Broderick} \& {Blandford}}{{Broderick} \&
  {Blandford}}{2004}]{bb:2004}
{Broderick} A.,  {Blandford} R.,  2004, \mn@doi [\mnras]
  {10.1111/j.1365-2966.2004.07582.x}, \href
  {http://esoads.eso.org/abs/2004MNRAS.349..994B} {349, 994}

\bibitem[\protect\citeauthoryear{{Broderick} \& {Loeb}}{{Broderick} \&
  {Loeb}}{2005}]{broderick:2005}
{Broderick} A.~E.,  {Loeb} A.,  2005, \mn@doi [\mnras]
  {10.1111/j.1365-2966.2005.09458.x}, \href
  {http://esoads.eso.org/abs/2005MNRAS.363..353B} {363, 353}

\bibitem[\protect\citeauthoryear{{Broderick} \& {Loeb}}{{Broderick} \&
  {Loeb}}{2009a}]{broderick:2009b}
{Broderick} A.~E.,  {Loeb} A.,  2009a, \mn@doi [\apj]
  {10.1088/0004-637X/697/2/1164}, \href
  {http://esoads.eso.org/abs/2009ApJ...697.1164B} {697, 1164}

\bibitem[\protect\citeauthoryear{{Broderick} \& {Loeb}}{{Broderick} \&
  {Loeb}}{2009b}]{broderickloeb:2009}
{Broderick} A.~E.,  {Loeb} A.,  2009b, \mn@doi [\apj]
  {10.1088/0004-637X/697/2/1164}, \href
  {http://adsabs.harvard.edu/abs/2009ApJ...697.1164B} {697, 1164}

\bibitem[\protect\citeauthoryear{{Broderick}, {Fish}, {Doeleman}  \&
  {Loeb}}{{Broderick} et~al.}{2009}]{broderick:2009a}
{Broderick} A.~E.,  {Fish} V.~L.,  {Doeleman} S.~S.,   {Loeb} A.,  2009,
  \mn@doi [\apj] {10.1088/0004-637X/697/1/45}, \href
  {http://esoads.eso.org/abs/2009ApJ...697...45B} {697, 45}

\bibitem[\protect\citeauthoryear{{Broderick}, {Fish}, {Doeleman}  \&
  {Loeb}}{{Broderick} et~al.}{2011a}]{broderick:2011a}
{Broderick} A.~E.,  {Fish} V.~L.,  {Doeleman} S.~S.,   {Loeb} A.,  2011a,
  \mn@doi [\apj] {10.1088/0004-637X/735/2/110}, \href
  {http://esoads.eso.org/abs/2011ApJ...735..110B} {735, 110}

\bibitem[\protect\citeauthoryear{{Broderick}, {Fish}, {Doeleman}  \&
  {Loeb}}{{Broderick} et~al.}{2011b}]{broderick:2011b}
{Broderick} A.~E.,  {Fish} V.~L.,  {Doeleman} S.~S.,   {Loeb} A.,  2011b,
  \mn@doi [\apj] {10.1088/0004-637X/738/1/38}, \href
  {http://esoads.eso.org/abs/2011ApJ...738...38B} {738, 38}

\bibitem[\protect\citeauthoryear{{Bromley}, {Melia}  \& {Liu}}{{Bromley}
  et~al.}{2001}]{bromley:2001}
{Bromley} B.~C.,  {Melia} F.,   {Liu} S.,  2001, \mn@doi [\apjl]
  {10.1086/322862}, \href {http://esoads.eso.org/abs/2001ApJ...555L..83B} {555,
  L83}

\bibitem[\protect\citeauthoryear{{Chan}, {Psaltis}  \& {{\"O}zel}}{{Chan}
  et~al.}{2013}]{chan:2013}
{Chan} C.-k.,  {Psaltis} D.,   {{\"O}zel} F.,  2013, \mn@doi [\apj]
  {10.1088/0004-637X/777/1/13}, \href
  {http://esoads.eso.org/abs/2013ApJ...777...13C} {777, 13}

\bibitem[\protect\citeauthoryear{{Chan}, {Psaltis}, {{\"O}zel}, {Narayan}  \&
  {Sadowski}}{{Chan} et~al.}{2015}]{chan:2015}
{Chan} C.-K.,  {Psaltis} D.,  {{\"O}zel} F.,  {Narayan} R.,   {Sadowski} A.,
  2015, \mn@doi [\apj] {10.1088/0004-637X/799/1/1}, \href
  {http://adsabs.harvard.edu/abs/2015ApJ...799....1C} {799, 1}

\bibitem[\protect\citeauthoryear{{Chan}, {Medeiros}, {Ozel}  \&
  {Psaltis}}{{Chan} et~al.}{2017}]{chan:2017}
{Chan} C.-k.,  {Medeiros} L.,  {Ozel} F.,   {Psaltis} D.,  2017, preprint,
  \href {http://esoads.eso.org/abs/2017arXiv170607062C} {} (\mn@eprint {arXiv}
  {1706.07062})

\bibitem[\protect\citeauthoryear{{Connors} \& {Stark}}{{Connors} \&
  {Stark}}{1977}]{connors:1977}
{Connors} P.~A.,  {Stark} R.~F.,  1977, \mn@doi [\nat] {10.1038/269128a0},
  \href {http://esoads.eso.org/abs/1977Natur.269..128C} {269, 128}

\bibitem[\protect\citeauthoryear{{Connors}, {Stark}  \& {Piran}}{{Connors}
  et~al.}{1980}]{connors:1980}
{Connors} P.~A.,  {Stark} R.~F.,   {Piran} T.,  1980, \mn@doi [\apj]
  {10.1086/157627}, \href {http://esoads.eso.org/abs/1980ApJ...235..224C} {235,
  224}

\bibitem[\protect\citeauthoryear{{Cunningham}}{{Cunningham}}{1975}]{cunningham:1975}
{Cunningham} C.~T.,  1975, \mn@doi [\apj] {10.1086/154033}, \href
  {http://esoads.eso.org/abs/1975ApJ...202..788C} {202, 788}

\bibitem[\protect\citeauthoryear{{Cunningham} \& {Bardeen}}{{Cunningham} \&
  {Bardeen}}{1973}]{cunningham:1973}
{Cunningham} C.~T.,  {Bardeen} J.~M.,  1973, \mn@doi [\apj] {10.1086/152223},
  \href {http://esoads.eso.org/abs/1973ApJ...183..237C} {183, 237}

\bibitem[\protect\citeauthoryear{{Dexter}}{{Dexter}}{2016}]{dexter:2016}
{Dexter} J.,  2016, \mn@doi [\mnras] {10.1093/mnras/stw1526}, \href
  {http://adsabs.harvard.edu/abs/2016MNRAS.462..115D} {462, 115}

\bibitem[\protect\citeauthoryear{{Dexter} \& {Agol}}{{Dexter} \&
  {Agol}}{2009}]{dexter:2009}
{Dexter} J.,  {Agol} E.,  2009, \mn@doi [\apj] {10.1088/0004-637X/696/2/1616},
  \href {http://esoads.eso.org/abs/2009ApJ...696.1616D} {696, 1616}

\bibitem[\protect\citeauthoryear{{Dexter} \& {Fragile}}{{Dexter} \&
  {Fragile}}{2011}]{dexter:2011}
{Dexter} J.,  {Fragile} P.~C.,  2011, \mn@doi [\apj]
  {10.1088/0004-637X/730/1/36}, \href
  {http://esoads.eso.org/abs/2011ApJ...730...36D} {730, 36}

\bibitem[\protect\citeauthoryear{{Dexter} \& {Fragile}}{{Dexter} \&
  {Fragile}}{2013}]{dexter:2013}
{Dexter} J.,  {Fragile} P.~C.,  2013, \mn@doi [\mnras] {10.1093/mnras/stt583},
  \href {http://esoads.eso.org/abs/2013MNRAS.432.2252D} {432, 2252}

\bibitem[\protect\citeauthoryear{{Dexter}, {Agol}, {Fragile}  \&
  {McKinney}}{{Dexter} et~al.}{2010}]{dexter:2010}
{Dexter} J.,  {Agol} E.,  {Fragile} P.~C.,   {McKinney} J.~C.,  2010, \mn@doi
  [\apj] {10.1088/0004-637X/717/2/1092}, \href
  {http://esoads.eso.org/abs/2010ApJ...717.1092D} {717, 1092}

\bibitem[\protect\citeauthoryear{{Dexter}, {McKinney}  \& {Agol}}{{Dexter}
  et~al.}{2012}]{dexter:2012}
{Dexter} J.,  {McKinney} J.~C.,   {Agol} E.,  2012, \mn@doi [\mnras]
  {10.1111/j.1365-2966.2012.20409.x}, \href
  {http://esoads.eso.org/abs/2012MNRAS.421.1517D} {421, 1517}

\bibitem[\protect\citeauthoryear{{Dolence}, {Gammie}, {Shiokawa}  \&
  {Noble}}{{Dolence} et~al.}{2012}]{dolence:2012}
{Dolence} J.~C.,  {Gammie} C.~F.,  {Shiokawa} H.,   {Noble} S.~C.,  2012,
  \mn@doi [\apjl] {10.1088/2041-8205/746/1/L10}, \href
  {http://esoads.eso.org/abs/2012ApJ...746L..10D} {746, L10}

\bibitem[\protect\citeauthoryear{{Eckart} et~al.,}{{Eckart}
  et~al.}{2008}]{eckart:2008}
{Eckart} A.,  et~al., 2008, \mn@doi [\aap] {10.1051/0004-6361:20078793}, \href
  {http://esoads.eso.org/abs/2008A%26A...479..625E} {479, 625}

\bibitem[\protect\citeauthoryear{{Falcke}, {Melia}  \& {Agol}}{{Falcke}
  et~al.}{2000}]{falcke:2000a}
{Falcke} H.,  {Melia} F.,   {Agol} E.,  2000, \mn@doi [\apjl] {10.1086/312423},
  \href {http://esoads.eso.org/abs/2000ApJ...528L..13F} {528, L13}

\bibitem[\protect\citeauthoryear{{Fraga-Encinas}, {Mo{\'s}cibrodzka},
  {Brinkerink}  \& {Falcke}}{{Fraga-Encinas} et~al.}{2016}]{fraga:2016}
{Fraga-Encinas} R.,  {Mo{\'s}cibrodzka} M.,  {Brinkerink} C.,   {Falcke} H.,
  2016, \mn@doi [\aap] {10.1051/0004-6361/201527599}, \href
  {http://esoads.eso.org/abs/2016A%26A...588A..57F} {588, A57}

\bibitem[\protect\citeauthoryear{{Gammie} \& {Leung}}{{Gammie} \&
  {Leung}}{2012}]{gammie:2012}
{Gammie} C.~F.,  {Leung} P.~K.,  2012, \mn@doi [\apj]
  {10.1088/0004-637X/752/2/123}, \href
  {http://esoads.eso.org/abs/2012ApJ...752..123G} {752, 123}

\bibitem[\protect\citeauthoryear{{Gammie}, {McKinney}  \& {T{\'o}th}}{{Gammie}
  et~al.}{2003}]{gammie:2003}
{Gammie} C.~F.,  {McKinney} J.~C.,   {T{\'o}th} G.,  2003, \mn@doi [\apj]
  {10.1086/374594}, \href {http://adsabs.harvard.edu/abs/2003ApJ...589..444G}
  {589, 444}

\bibitem[\protect\citeauthoryear{{Gold}, {McKinney}, {Johnson}  \&
  {Doeleman}}{{Gold} et~al.}{2016}]{gold:2016}
{Gold} R.,  {McKinney} J.~C.,  {Johnson} M.~D.,   {Doeleman} S.~S.,  2016,
  preprint, \href {http://adsabs.harvard.edu/abs/2016arXiv160105550G} {}
  (\mn@eprint {arXiv} {1601.05550})

\bibitem[\protect\citeauthoryear{{Hamaker} \& {Bregman}}{{Hamaker} \&
  {Bregman}}{1996}]{hamaker:1996}
{Hamaker} J.~P.,  {Bregman} J.~D.,  1996, \aaps, \href
  {http://adsabs.harvard.edu/abs/1996A%26AS..117..161H} {117, 161}

\bibitem[\protect\citeauthoryear{{Huang}, {Liu}, {Shen}, {Cai}, {Li}  \&
  {Fryer}}{{Huang} et~al.}{2008}]{huang:2008}
{Huang} L.,  {Liu} S.,  {Shen} Z.-Q.,  {Cai} M.~J.,  {Li} H.,   {Fryer} C.~L.,
  2008, \mn@doi [\apjl] {10.1086/587742}, \href
  {http://esoads.eso.org/abs/2008ApJ...676L.119H} {676, L119}

\bibitem[\protect\citeauthoryear{{Ishihara}, {Takahashi}  \&
  {Tomimatsu}}{{Ishihara} et~al.}{1988}]{ishihara:1988}
{Ishihara} H.,  {Takahashi} M.,   {Tomimatsu} A.,  1988, \mn@doi [\prd]
  {10.1103/PhysRevD.38.472}, \href
  {http://adsabs.harvard.edu/abs/1988PhRvD..38..472I} {38, 472}

\bibitem[\protect\citeauthoryear{{Janett}, {Carlin}, {Steiner}  \&
  {Belluzzi}}{{Janett} et~al.}{2017}]{janett:2017}
{Janett} G.,  {Carlin} E.~S.,  {Steiner} O.,   {Belluzzi} L.,  2017, \mn@doi
  [\apj] {10.3847/1538-4357/aa671d}, \href
  {http://adsabs.harvard.edu/abs/2017ApJ...840..107J} {840, 107}

\bibitem[\protect\citeauthoryear{{Johnson} et~al.,}{{Johnson}
  et~al.}{2015}]{johnson:2015}
{Johnson} M.~D.,  et~al., 2015, \mn@doi [Science] {10.1126/science.aac7087},
  \href {http://esoads.eso.org/abs/2015Sci...350.1242J} {350, 1242}

\bibitem[\protect\citeauthoryear{{Kosowsky}}{{Kosowsky}}{1996}]{kosowsky:1996}
{Kosowsky} A.,  1996, \mn@doi [Annals of Physics] {10.1006/aphy.1996.0020},
  \href {http://adsabs.harvard.edu/abs/1996AnPhy.246...49K} {246, 49}

\bibitem[\protect\citeauthoryear{{Kuo} et~al.,}{{Kuo} et~al.}{2014}]{kuo:2014}
{Kuo} C.~Y.,  et~al., 2014, \mn@doi [\apjl] {10.1088/2041-8205/783/2/L33},
  \href {http://adsabs.harvard.edu/abs/2014ApJ...783L..33K} {783, L33}

\bibitem[\protect\citeauthoryear{{Landi Degl'Innocenti} \& {Landi
  Degl'Innocenti}}{{Landi Degl'Innocenti} \& {Landi
  Degl'Innocenti}}{1985}]{LDI:1985}
{Landi Degl'Innocenti} E.,  {Landi Degl'Innocenti} M.,  1985, \mn@doi
  [\solphys] {10.1007/BF00165988}, \href
  {http://esoads.eso.org/abs/1985SoPh...97..239L} {97, 239}

\bibitem[\protect\citeauthoryear{{L{\'o}pez Ariste} \& {Semel}}{{L{\'o}pez
  Ariste} \& {Semel}}{1999}]{lopezariste:1999}
{L{\'o}pez Ariste} A.,  {Semel} M.,  1999, \aap, \href
  {http://esoads.eso.org/abs/1999A%26A...350.1089L} {350, 1089}

\bibitem[\protect\citeauthoryear{{Mao}, {Dexter}  \& {Quataert}}{{Mao}
  et~al.}{2017}]{mao:2017}
{Mao} S.~A.,  {Dexter} J.,   {Quataert} E.,  2017, \mn@doi [\mnras]
  {10.1093/mnras/stw3324}, \href
  {http://esoads.eso.org/abs/2017MNRAS.466.4307M} {466, 4307}

\bibitem[\protect\citeauthoryear{{Marrone}, {Moran}, {Zhao}  \&
  {Rao}}{{Marrone} et~al.}{2007}]{marrone:2007}
{Marrone} D.~P.,  {Moran} J.~M.,  {Zhao} J.-H.,   {Rao} R.,  2007, \mn@doi
  [\apjl] {10.1086/510850}, \href
  {http://esoads.eso.org/abs/2007ApJ...654L..57M} {654, L57}

\bibitem[\protect\citeauthoryear{{Marrone} et~al.,}{{Marrone}
  et~al.}{2008}]{marrone:2008}
{Marrone} D.~P.,  et~al., 2008, \mn@doi [\apj] {10.1086/588806}, \href
  {http://esoads.eso.org/abs/2008ApJ...682..373M} {682, 373}

\bibitem[\protect\citeauthoryear{{Medeiros}, {Chan}, {{\"O}zel}, {Psaltis},
  {Kim}, {Marrone}  \& {Sadowski}}{{Medeiros} et~al.}{2017}]{medeiros:2017}
{Medeiros} L.,  {Chan} C.-k.,  {{\"O}zel} F.,  {Psaltis} D.,  {Kim} J.,
  {Marrone} D.~P.,   {Sadowski} A.,  2017, \mn@doi [\apj]
  {10.3847/1538-4357/aa7751}, \href
  {http://esoads.eso.org/abs/2017ApJ...844...35M} {844, 35}

\bibitem[\protect\citeauthoryear{{Mo{\'s}cibrodzka}, {Gammie}, {Dolence},
  {Shiokawa}  \& {Leung}}{{Mo{\'s}cibrodzka} et~al.}{2009}]{moscibrodzka:2009}
{Mo{\'s}cibrodzka} M.,  {Gammie} C.~F.,  {Dolence} J.~C.,  {Shiokawa} H.,
  {Leung} P.~K.,  2009, \mn@doi [\apj] {10.1088/0004-637X/706/1/497}, \href
  {http://esoads.eso.org/abs/2009ApJ...706..497M} {706, 497}

\bibitem[\protect\citeauthoryear{{Mo{\'s}cibrodzka}, {Shiokawa}, {Gammie}  \&
  {Dolence}}{{Mo{\'s}cibrodzka} et~al.}{2012}]{moscibrodzka:2012}
{Mo{\'s}cibrodzka} M.,  {Shiokawa} H.,  {Gammie} C.~F.,   {Dolence} J.~C.,
  2012, \mn@doi [\apjl] {10.1088/2041-8205/752/1/L1}, \href
  {http://esoads.eso.org/abs/2012ApJ...752L...1M} {752, L1}

\bibitem[\protect\citeauthoryear{{Mo{\'s}cibrodzka}, {Falcke}, {Shiokawa}  \&
  {Gammie}}{{Mo{\'s}cibrodzka} et~al.}{2014}]{moscibrodzka:2014}
{Mo{\'s}cibrodzka} M.,  {Falcke} H.,  {Shiokawa} H.,   {Gammie} C.~F.,  2014,
  \mn@doi [\aap] {10.1051/0004-6361/201424358}, \href
  {http://adsabs.harvard.edu/abs/2014A%26A...570A...7M} {570, A7}

\bibitem[\protect\citeauthoryear{{Mo{\'s}cibrodzka}, {Falcke}  \&
  {Shiokawa}}{{Mo{\'s}cibrodzka} et~al.}{2016}]{moscibrodzka:2016}
{Mo{\'s}cibrodzka} M.,  {Falcke} H.,   {Shiokawa} H.,  2016, \mn@doi [\aap]
  {10.1051/0004-6361/201526630}, \href
  {http://adsabs.harvard.edu/abs/2016A%26A...586A..38M} {586, A38}

\bibitem[\protect\citeauthoryear{{Mo{\'s}cibrodzka}, {Dexter}, {Davelaar}  \&
  {Falcke}}{{Mo{\'s}cibrodzka} et~al.}{2017}]{moscibrodzka:2017}
{Mo{\'s}cibrodzka} M.,  {Dexter} J.,  {Davelaar} J.,   {Falcke} H.,  2017,
  \mn@doi [\mnras] {10.1093/mnras/stx587}, \href
  {http://esoads.eso.org/abs/2017MNRAS.468.2214M} {468, 2214}

\bibitem[\protect\citeauthoryear{{Mu{\~n}oz}, {Marrone}  \&
  {Moran}}{{Mu{\~n}oz} et~al.}{2009}]{munoz:2009}
{Mu{\~n}oz} D.,  {Marrone} D.,   {Moran} J.,  2009, in American Astronomical
  Society Meeting Abstracts \#214. p.~761

\bibitem[\protect\citeauthoryear{{Mu{\~n}oz}, {Marrone}, {Moran}  \&
  {Rao}}{{Mu{\~n}oz} et~al.}{2012}]{munoz:2012}
{Mu{\~n}oz} D.~J.,  {Marrone} D.~P.,  {Moran} J.~M.,   {Rao} R.,  2012, \mn@doi
  [\apj] {10.1088/0004-637X/745/2/115}, \href
  {http://esoads.eso.org/abs/2012ApJ...745..115M} {745, 115}

\bibitem[\protect\citeauthoryear{{Noble}, {Gammie}, {McKinney}  \& {Del
  Zanna}}{{Noble} et~al.}{2006}]{noble:2006}
{Noble} S.~C.,  {Gammie} C.~F.,  {McKinney} J.~C.,   {Del Zanna} L.,  2006,
  \mn@doi [\apj] {10.1086/500349}, \href
  {http://adsabs.harvard.edu/abs/2006ApJ...641..626N} {641, 626}

\bibitem[\protect\citeauthoryear{{Noble}, {Leung}, {Gammie}  \& {Book}}{{Noble}
  et~al.}{2007}]{noble:2007}
{Noble} S.~C.,  {Leung} P.~K.,  {Gammie} C.~F.,   {Book} L.~G.,  2007, \mn@doi
  [Classical and Quantum Gravity] {10.1088/0264-9381/24/12/S17}, \href
  {http://adsabs.harvard.edu/abs/2007CQGra..24S.259N} {24, S259}

\bibitem[\protect\citeauthoryear{{Peraiah}}{{Peraiah}}{2001}]{peraiah:2001}
{Peraiah} A.,  2001, {An Introduction to Radiative Transfer}.
Cambridge University Press

\bibitem[\protect\citeauthoryear{{Porth}, {Olivares}, {Mizuno}, {Younsi},
  {Rezzolla}, {Moscibrodzka}, {Falcke}  \& {Kramer}}{{Porth}
  et~al.}{2017}]{porth:2017}
{Porth} O.,  {Olivares} H.,  {Mizuno} Y.,  {Younsi} Z.,  {Rezzolla} L.,
  {Moscibrodzka} M.,  {Falcke} H.,   {Kramer} M.,  2017, \mn@doi [Computational
  Astrophysics and Cosmology] {10.1186/s40668-017-0020-2}, \href
  {http://esoads.eso.org/abs/2017ComAC...4....1P} {4, 1}

\bibitem[\protect\citeauthoryear{{Pu}, {Yun}, {Younsi}  \& {Yoon}}{{Pu}
  et~al.}{2016a}]{pu:2016a}
{Pu} H.-Y.,  {Yun} K.,  {Younsi} Z.,   {Yoon} S.-J.,  2016a, \mn@doi [\apj]
  {10.3847/0004-637X/820/2/105}, \href
  {http://esoads.eso.org/abs/2016ApJ...820..105P} {820, 105}

\bibitem[\protect\citeauthoryear{{Pu}, {Akiyama}  \& {Asada}}{{Pu}
  et~al.}{2016b}]{pu:2016b}
{Pu} H.-Y.,  {Akiyama} K.,   {Asada} K.,  2016b, \mn@doi [\apj]
  {10.3847/0004-637X/831/1/4}, \href
  {http://esoads.eso.org/abs/2016ApJ...831....4P} {831, 4}

\bibitem[\protect\citeauthoryear{{Rees}, {Durrant}  \& {Murphy}}{{Rees}
  et~al.}{1989}]{rees:1989}
{Rees} D.~E.,  {Durrant} C.~J.,   {Murphy} G.~A.,  1989, \mn@doi [\apj]
  {10.1086/167364}, \href {http://adsabs.harvard.edu/abs/1989ApJ...339.1093R}
  {339, 1093}

\bibitem[\protect\citeauthoryear{{Roelofs}, {Johnson}, {Shiokawa}, {Doeleman}
  \& {Falcke}}{{Roelofs} et~al.}{2017}]{roelofs:2017}
{Roelofs} F.,  {Johnson} M.~D.,  {Shiokawa} H.,  {Doeleman} S.~S.,   {Falcke}
  H.,  2017, preprint, \href {http://esoads.eso.org/abs/2017arXiv170801056R} {}
  (\mn@eprint {arXiv} {1708.01056})

\bibitem[\protect\citeauthoryear{{Sereno}}{{Sereno}}{2005}]{sereno:2005}
{Sereno} M.,  2005, \mn@doi [\mnras] {10.1111/j.1365-2966.2004.08456.x}, \href
  {http://esoads.eso.org/abs/2005MNRAS.356..381S} {356, 381}

\bibitem[\protect\citeauthoryear{{Shahzamanian} et~al.,}{{Shahzamanian}
  et~al.}{2015}]{sha:2015}
{Shahzamanian} B.,  et~al., 2015, \mn@doi [\aap] {10.1051/0004-6361/201425239},
  \href {http://esoads.eso.org/abs/2015A%26A...576A..20S} {576, A20}

\bibitem[\protect\citeauthoryear{{Shcherbakov}}{{Shcherbakov}}{2008}]{shcherbakov:2008}
{Shcherbakov} R.~V.,  2008, \mn@doi [\apj] {10.1086/592326}, \href
  {http://adsabs.harvard.edu/abs/2008ApJ...688..695S} {688, 695}

\bibitem[\protect\citeauthoryear{{Shcherbakov}, {Penna}  \&
  {McKinney}}{{Shcherbakov} et~al.}{2012}]{romans:2012}
{Shcherbakov} R.~V.,  {Penna} R.~F.,   {McKinney} J.~C.,  2012, \mn@doi [\apj]
  {10.1088/0004-637X/755/2/133}, \href
  {http://adsabs.harvard.edu/abs/2012ApJ...755..133S} {755, 133}

\bibitem[\protect\citeauthoryear{{Shiokawa}, {Gammie}  \&
  {Doeleman}}{{Shiokawa} et~al.}{2017}]{shiokawa:2017}
{Shiokawa} H.,  {Gammie} C.~F.,   {Doeleman} S.~S.,  2017, preprint, \href
  {http://esoads.eso.org/abs/2017arXiv170802577S} {} (\mn@eprint {arXiv}
  {1708.02577})

\bibitem[\protect\citeauthoryear{{Vincent}, {Paumard}, {Gourgoulhon}  \&
  {Perrin}}{{Vincent} et~al.}{2011}]{vincent:2011}
{Vincent} F.~H.,  {Paumard} T.,  {Gourgoulhon} E.,   {Perrin} G.,  2011,
  \mn@doi [Classical and Quantum Gravity] {10.1088/0264-9381/28/22/225011},
  \href {http://esoads.eso.org/abs/2011CQGra..28v5011V} {28, 225011}

\bibitem[\protect\citeauthoryear{{Vincent}, {Yan}, {Straub}, {Zdziarski}  \&
  {Abramowicz}}{{Vincent} et~al.}{2015}]{vincent:2015}
{Vincent} F.~H.,  {Yan} W.,  {Straub} O.,  {Zdziarski} A.~A.,   {Abramowicz}
  M.~A.,  2015, \mn@doi [\aap] {10.1051/0004-6361/201424306}, \href
  {http://esoads.eso.org/abs/2015A%26A...574A..48V} {574, A48}

\bibitem[\protect\citeauthoryear{{Wang} \& {Bovik}}{{Wang} \&
  {Bovik}}{2002}]{wang:2002}
{Wang} Z.,  {Bovik} A.~C.,  2002, \mn@doi [IEEE Signal Processing Letters]
  {10.1109/97.995823}, \href {http://ieeexplore.ieee.org/document/995823/} {9,
  81}

\bibitem[\protect\citeauthoryear{{Weinberg}}{{Weinberg}}{2008}]{weinberg:2008}
{Weinberg} S.,  2008, {Cosmology}.
Oxford University Press

\bibitem[\protect\citeauthoryear{{White}, {Stone}  \& {Gammie}}{{White}
  et~al.}{2016}]{white:2016}
{White} C.~J.,  {Stone} J.~M.,   {Gammie} C.~F.,  2016, \mn@doi [\apjs]
  {10.3847/0067-0049/225/2/22}, \href
  {http://esoads.eso.org/abs/2016ApJS..225...22W} {225, 22}

\bibitem[\protect\citeauthoryear{{Younsi} \& {Wu}}{{Younsi} \&
  {Wu}}{2015}]{younsi:2015}
{Younsi} Z.,  {Wu} K.,  2015, \mn@doi [\mnras] {10.1093/mnras/stv2203}, \href
  {http://esoads.eso.org/abs/2015MNRAS.454.3283Y} {454, 3283}

\bibitem[\protect\citeauthoryear{{Younsi}, {Wu}  \& {Fuerst}}{{Younsi}
  et~al.}{2012}]{younsi:2012}
{Younsi} Z.,  {Wu} K.,   {Fuerst} S.~V.,  2012, \mn@doi [\aap]
  {10.1051/0004-6361/201219599}, \href
  {http://esoads.eso.org/abs/2012A%26A...545A..13Y} {545, A13}

\bibitem[\protect\citeauthoryear{{Yuan}, {Cao}, {Huang}  \& {Shen}}{{Yuan}
  et~al.}{2009}]{yuan:2009}
{Yuan} Y.-F.,  {Cao} X.,  {Huang} L.,   {Shen} Z.-Q.,  2009, \mn@doi [\apj]
  {10.1088/0004-637X/699/1/722}, \href
  {http://esoads.eso.org/abs/2009ApJ...699..722Y} {699, 722}

\bibitem[\protect\citeauthoryear{{van Ballegooijen}}{{van
  Ballegooijen}}{1985}]{vanball:1985}
{van Ballegooijen} A.~A.,  1985, in {Hagyard} M.~J.,  ed., Measurements of
  Solar Vector Magnetic Fields.

\makeatother
\end{thebibliography}



\appendix
\section{Special solutions to polarized transfer equation}

It may be useful for tests to have simplified analytic solutions to the polarized transfer equation (\ref{spoltrans}) in special cases.  Here we consider solutions with Faraday rotation alone (and no absorption and emission), and when Faraday rotation is absent.

\subsection{Solution with Faraday rotation alone}

Consider (\ref{spoltrans}) with the only the rotation coefficients nonzero:
\begin{equation}
\frac{d}{d\lambda}
\begin{pmatrix} I \\ Q \\ U \\ V \end{pmatrix}
= -\begin{pmatrix} 
0 & 0 & 0 & 0 \\
0 & 0 & \rho_V & -\rho_U \\
0 & -\rho_V & 0 & \rho_Q \\
0 & \rho_U & -\rho_Q & 0 \\
\end{pmatrix}
\begin{pmatrix} I \\ Q \\ U \\ V \end{pmatrix}.
\end{equation}
This can be integrated directly to find the analytic solution:
\begin{equation}
I = I_0
\end{equation}
\begin{equation}
\begin{split}
Q = Q_0 \cos(\rho \lambda) & + 2 \, \frac{\rho_Q (\bbrho \cdot \bS)}{\rho^2} \sin^2(\rho \lambda/2) \\
	& \qquad\qquad + \frac{\rho_U V_0 - \rho_V U_0}{\rho} \sin(\rho \lambda)
\end{split},
\end{equation}
\begin{equation}
\begin{split}
U = U_0 \cos(\rho \lambda) & + 2 \, \frac{\rho_U (\bbrho \cdot \bS)}{\rho^2} \sin^2(\rho \lambda/2) \\
	& \qquad\qquad + \frac{\rho_V Q_0 - \rho_Q V_0}{\rho} \sin(\rho \lambda)
\end{split},
\end{equation}
\begin{equation}
\begin{split}
V = V_0 \cos(\rho \lambda) & + 2 \, \frac{\rho_V (\bbrho \cdot \bS)}{\rho^2} \sin^2(\rho \lambda/2) \\
	& \qquad\qquad + \frac{\rho_Q U_0 - \rho_U Q_0}{\rho} \sin(\rho \lambda),
\end{split}
\end{equation}
which has a pleasing symmetry to it.  Here $\rho^2 \equiv \rho_Q^2 + \rho_U^2 + \rho_V^2$, and 
$\bbrho \cdot \bS \equiv \rho_Q Q_0 + \rho_U U_0 + \rho_V V_0$. 

\subsection{Solution with emission and absorption alone}

Now consider the piece of (\ref{spoltrans}) with $\rho_A \rightarrow 0$: 
\begin{equation}
\frac{d}{d\lambda}
\begin{pmatrix} I \\ Q \\ U \\ V \end{pmatrix}
 = \begin{pmatrix} j_{I} \\ j_{Q} \\ j_{U} \\ j_{V} \end{pmatrix}
- \begin{pmatrix} 
\alpha_{I} & \alpha_{Q} & \alpha_{U} & \alpha_{V} \\
\alpha_{Q} & \alpha_{I} & 0 & 0 \\
\alpha_{U} & 0 & \alpha_{I} & 0 \\
\alpha_{V} & 0 & 0 & \alpha_{I}  
\end{pmatrix}
\begin{pmatrix} I \\ Q \\ U \\ V \end{pmatrix}.
\end{equation}
The matrix $\bM$ on the RHS is real and symmetric, so one can solve by finding the
eigenvalues and eigenvectors of $\bM$, projecting the initial state and
emission coefficients into the eigenbasis, where (\ref{spoltrans}) reduces to
the same form as the unpolarized radiative transfer equation, and reassembling
the result in the Stokes basis. 

Rather than simply stating the result, it may be helpful to give a few intermediate
results.  Here is an orthonormal eigenbasis for $\bM$ (in the Stokes basis):
\begin{equation}
e_1 = \frac{1}{N_1}\{0, -\alpha_V n_U, -\alpha_V n_Q, \alpha_U n_Q + \alpha_Q n_U \},
\end{equation}
with eigenvalue $1/\lambda_1 = \alpha_I$,
$N_1 \equiv (2 (\alpha_Q \alpha_U n_Q n_U + n_Q^2 n_U^2))^{1/2}$,
$n_Q \equiv (\alpha_Q^2 + \alpha_V^2)^{1/2}$, 
$n_U \equiv (\alpha_U^2 + \alpha_V^2)^{1/2}$,
\begin{equation}
e_2 = \frac{1}{N_2}\{0,  \alpha_V n_U, -\alpha_V n_Q, \alpha_U n_Q - \alpha_Q n_U \}/N_2,
\end{equation}
with eigenvalue $1/\lambda_2 = \alpha_I$,
$N_2 \equiv (2 (-\alpha_Q \alpha_U n_Q n_U + n_Q^2 n_U^2))^{1/2}$,
\begin{equation}
e_3 = \frac{1}{\sqrt{2}\alpha_P}\{\alpha_P, \alpha_Q, \alpha_U, \alpha_V\},
\end{equation}
with eigenvalue $1/\lambda_3 = \alpha_I + \alpha_P$, 
$\alpha_P \equiv (\alpha_Q^2 + \alpha_U^2 + \alpha_V^2)^{1/2}$,
\begin{equation}
e_4 = \frac{1}{\sqrt{2}\alpha_P} \{-\alpha_P, \alpha_Q, \alpha_U, \alpha_V\},
\end{equation}
with eigenvalue $1/\lambda_4 = \alpha_I - \alpha_P$.  Evidently if $I$ is to decay under absorption we must have $\alpha_I \ge \alpha_P$. Notice that $1/\lambda_i$ is an eigenvalue, and $\lambda$ is the
affine parameter. 

Finding the combined absorption and emission solution is now easy. Let $a_i(\lambda)$ be the solution for the amplitude of eigenvector $e_i$.   The transfer equation in the eigenbasis, excluding Faraday conversion, is
\begin{equation}
\frac{d a_i}{d\lambda} = j_i - \frac{a_i}{\lambda_i}
\end{equation}
where $j_i = j_A e_{i,A}$.  The solution is identical to the formal solution of the unpolarized transfer equation:
\begin{equation}
a_i(\lambda) = j_i \lambda_i (1 - e^{- \lambda/ \lambda_i}) 
+ a^0_i e^{- \lambda / \lambda_i}.
\end{equation}
Here $a_i^0$ is the initial Stokes vector projected into the eigenbasis.  The solution in the Stokes basis is then
\begin{equation}
S_A(\lambda) = a_i(\lambda) e_{i,A}.
\end{equation}

The final result can be written:
\begin{equation}\label{eq:EAI}
\begin{aligned}
I = &\left( I_0 \chaPs - \frac{\acbS}{\aP} \shaPs \right) \eaIs \\
&+ \frac{\acj}{\aI^2- \aP^2} \left(-1 + \frac{\aI \shaPs  + \aP
  \chaPs}{\aP} \eaPs \right) \\
&+ \frac{\aI \jI }{\aI^2-\aP^2} \left(1 -
\frac{\aI \chaPs + \aP \shaPs}{\aI}\eaPs \right),
\end{aligned}
\end{equation}
\begin{equation}\label{eq:EAQ}
\begin{aligned}
Q = & \left( Q_0  +  \frac{\aQ \acbS}{\aP^2} (\chaPs -1) - I_0 \frac{\aQ}{\aP} \shaPs \right) \eaIs\\
        & +  \frac{\jQ(1 - \eaIs)}{\aI}\\
        & + \frac{(\acj)\aQ}{\aI (\aI^2 - \aP^2)} \\
        & \left(1 - [(1 - \frac{\aI^2}{\aP^2}) - \frac{\aI}{\aP^2} (\aI  \chaPs + \aP \shaPs )]  \eaIs \right) \\
        & + \frac{\jI \aQ}{\aP(\aI^2 - \aP^2)} \\
        & \left( -\aP + (\aP  \chaPs + \aI \shaPs )  \eaIs \right),
\end{aligned}
\end{equation}
\begin{equation}\label{eq:EAU}
\begin{aligned}
U = & \left( U_0  +  \frac{\aU \acbS}{\aP^2} (\chaPs -1) - I_0 \frac{\aU}{\aP} \shaPs \right) \eaIs\\
        & +  \frac{\jU(1 - \eaIs)}{\aI}\\
        & + \frac{(\acj)\aU}{\aI (\aI^2 - \aP^2)} \\
        & \left(1 - [(1 - \frac{\aI^2}{\aP^2}) - \frac{\aI}{\aP^2} (\aI  \chaPs + \aP \shaPs )]  \eaIs \right) \\
        & + \frac{\jI \aU}{\aP(\aI^2 - \aP^2)} \\
        & \left( -\aP + (\aP  \chaPs + \aI \shaPs )  \eaIs \right),
\end{aligned}
\end{equation}
\begin{equation}\label{eq:EAV}
\begin{aligned}
V = & \left( V_0  +  \frac{\aV \acbS}{\aP^2} (\chaPs -1) - I_0 \frac{\aV}{\aP} \shaPs \right) \eaIs\\
        & +  \frac{\jV(1 - \eaIs)}{\aI}\\
        & + \frac{(\acj)\aV}{\aI (\aI^2 - \aP^2)} \\
        & \left(1 - [(1 - \frac{\aI^2}{\aP^2}) - \frac{\aI}{\aP^2} (\aI  \chaPs + \aP \shaPs )]  \eaIs \right) \\
        & + \frac{\jI \aV}{\aP(\aI^2 - \aP^2)} \\
        & \left( -\aP + (\aP  \chaPs + \aI \shaPs )  \eaIs \right),
\end{aligned}
\end{equation}
where $\alpha_P^2 = \alpha_Q^2 + \alpha_U^2 + \alpha_V^2$, $\alpha \cdot \bS = \alpha_Q Q_0 + \alpha_U U_0 + \alpha_V V_0$, and $\alpha \cdot j= \alpha_Q j_Q + \alpha_U j_U + \alpha_V j_V $. If we ignore emission only the first terms in (\ref{eq:EAI})-(\ref{eq:EAV}) do not vanish. If $\alpha_P \rightarrow 0$ (or $\alpha_I \rightarrow 0$) then there
is a danger of division by zero and one must take the appropriate limit analytically.

The general solution is found in a similar way (see \LDI).
Because the matrix $K_{AB}$ is not symmetric, the eigenvalues
are complex, so there are both oscillatory and exponentially growing/decaying components to the solution.   


\bsp	
\label{lastpage}
\end{document}